\documentclass[12pt]{iopart}

\usepackage{graphicx}
\usepackage{subcaption}
\usepackage{orcidlink}

\usepackage{amsfonts}
\expandafter\let\csname equation*\endcsname\relax

\expandafter\let\csname endequation*\endcsname\relax

\usepackage{amsmath}
\usepackage{listings}
\usepackage{xcolor}

\usepackage{ulem}

\usepackage[most, breakable]{tcolorbox}
\usepackage{hyperref}

\hypersetup{
    breaklinks = true,
    colorlinks=true,
    linkcolor=blue,
    filecolor=magenta,      
    urlcolor=blue,
    citecolor=red
}

\definecolor{bg}{RGB}{255,249,227}
\definecolor{light_orange}{RGB}{255,193,134}
\newcommand{\red}[1]{#1}

\newtcolorbox{optional}[1][]{%
enhanced jigsaw,
colback=bg,
boxrule=0pt,
overlay unbroken and first ={%
\draw[line width=0.2pt,double=bg,draw=bg!70!black,
    double distance=1pt,] (frame.north west) -- (frame.north east);
\draw[line width=0.2pt,double=bg,draw=bg!70!black,
    double distance=1pt,] (frame.south west) -- (frame.south east);},
breakable,
arc=0pt,outer arc=0pt,
#1}%

\newtcolorbox{tldr}[1][]{%
enhanced jigsaw,
colback=light_orange,
boxrule=0pt,
overlay unbroken and first ={%
\draw[line width=0.2pt,double=bg,draw=bg!70!black,
    double distance=1pt,] (frame.north west) -- (frame.north east);
\draw[line width=0.2pt,double=bg,draw=bg!70!black,
    double distance=1pt,] (frame.south west) -- (frame.south east);},
arc=0pt,outer arc=0pt,
#1}%

\definecolor{codegreen}{rgb}{0,0.6,0}
\definecolor{codegray}{rgb}{0.5,0.5,0.5}
\definecolor{codepurple}{rgb}{0.58,0,0.82}
\definecolor{backcolour}{rgb}{0.95,0.95,0.92}

\lstdefinestyle{mystyle}{
    backgroundcolor=\color{backcolour},   
    commentstyle=\color{codegreen},
    keywordstyle=\color{magenta},
    numberstyle=\tiny\color{codegray},
    stringstyle=\color{codepurple},
    basicstyle=\ttfamily,
    breakatwhitespace=false,         
    breaklines=true,                 
    captionpos=b,                    
    keepspaces=true,                 
    numbers=left,                    
    numbersep=5pt,                  
    showspaces=false,                
    showstringspaces=false,
    showtabs=false,                  
    tabsize=2
}
\lstMakeShortInline[columns=fixed]!
\newcommand{\be}{\begin{equation}}
\newcommand{\ee}{\end{equation}}

\begin{document}

\title{A practical guide to Digital Micro-mirror Devices (DMDs) for wavefront shaping}

\author{Sébastien M. Popoff~\footnote{corresponding author}~\orcidlink{0000-0002-7199-9814}}
\address{Institut Langevin, ESPCI Paris, PSL University, CNRS, France}
\ead{sebastien.popoff@espci.psl.eu}

\author{Louis Malosse~\orcidlink{0009-0004-9006-9374}}
\address{LIRA, Observatoire de Paris, Université PSL, CNRS, Université Paris Cité, Sorbonne Université, France}
\address{ISMO, University Paris Saclay, CNRS, France}
\ead{louis.malosse@universite-paris-saclay.fr}

\author{Rodrigo Gutiérrez-Cuevas~\orcidlink{0000-0002-3451-6684}}
\address{Institut Langevin, ESPCI Paris, PSL University, CNRS, France}
\ead{rodrigo.gutierrez-cuevas@espci.psl.eu}

\author{Yaron Bromberg~\orcidlink{0000-0003-2565-7394}}
\address{Racah Institute of Physics, The Hebrew University of Jerusalem, Israel}

\author{Jean Commère~\orcidlink{0009-0000-4710-4972}}
\address{LIRA, Observatoire de Paris, Université PSL, CNRS, Université Paris Cité, Sorbonne Université, France}
\ead{jean.commere@obspm.fr}

\author{Marie Glanc}
\address{LIRA, Observatoire de Paris, Université PSL, CNRS, Université Paris Cité, Sorbonne Université, France}
\ead{marie.glanc@obspm.fr}

\author{Raphaël Galicher}
\address{LIRA, Observatoire de Paris, Université PSL, CNRS, Université Paris Cité, Sorbonne Université, France}
\ead{raphael.galicher@obspm.fr}

\author{Maxime W. Matthès}
\address{Institut Langevin, ESPCI Paris, PSL University, CNRS, France}

\vspace{10pt}
\begin{indented}
  \item[]September 2023
\end{indented}

\begin{abstract}
  Digital micromirror devices have gained popularity in wavefront shaping,
  offering a high frame rate alternative to liquid crystal spatial light modulators.
  They are relatively inexpensive, offer high resolution,
  are easy to operate, and a single device can be used in a broad optical bandwidth.
  However, some technical drawbacks must be considered
  to achieve optimal performance.
  These issues, often undocumented by manufacturers,
  mostly stem from the device's original design for video projection applications.
  Herein, we present a guide to characterize and mitigate these effects.
  Our focus is on providing simple and practical solutions
  that can be easily incorporated into a typical wavefront shaping setup.
\end{abstract}

%
%
%
%
%

\section{Introduction}

Since the advent of adaptive optics, various technologies have been employed
to modulate the amplitude and/or phase of light.
Early adaptive optics devices, utilized in fields like microscopy and astronomy,
offer rapid modulation capable of compensating for the aberrations of optical systems
in real-time.
However, these devices are constrained by a limited number of actuators,
restricting their utility in complex media where a large number of degrees of freedom is essential.
Liquid Crystal Spatial Light Modulators (LC-SLMs),
which allow for the control of light phase across typically more than a million pixels,
have emerged as powerful tools for wavefront shaping in complex media
since the seminal work of A. \red{Mosk} and I. Vellekoop in the mid-2000s~\cite{Vellekoop2007focusing}.
Nonetheless, LC-SLMs are hampered by their slow response time,
permitting only a modulation speed ranging from a few Hz to \red{a few hundred} Hz.\\

Digital Micromirror Devices (DMDs) have emerged as a technology bridging the gap
between these two types of systems;
they offer a large number of pixels (similar to LC-SLMs) and fast modulation speeds (typically up to several tens of kHz).
Their high speed capabilities made them attractive for real-time applications,
in particular for high-resolution imaging microscopy
requiring fast scanning or illumination shaping~\cite{cha2000nontranslational,Zhuang2020},
biolithography~\cite{yoon2018emerging},
and optical tweezers~\cite{gauthier2016direct}.
However, DMDs are restricted to hardware binary amplitude modulation and are not optimized for coherent light applications.
Utilizing DMDs for coherent control of light in complex media is therefore non-trivial
and necessitates specific adaptations for efficient use.\\

\begin{figure}
  \centering
  \begin{subfigure}{0.49\textwidth}
    \centering
    \includegraphics[width = \textwidth]{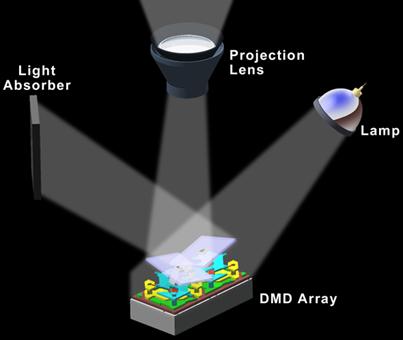}
    \label{fig:pix_left}
  \end{subfigure}
  \begin{subfigure}{0.49\textwidth}
    \centering
    \includegraphics[width = \textwidth]{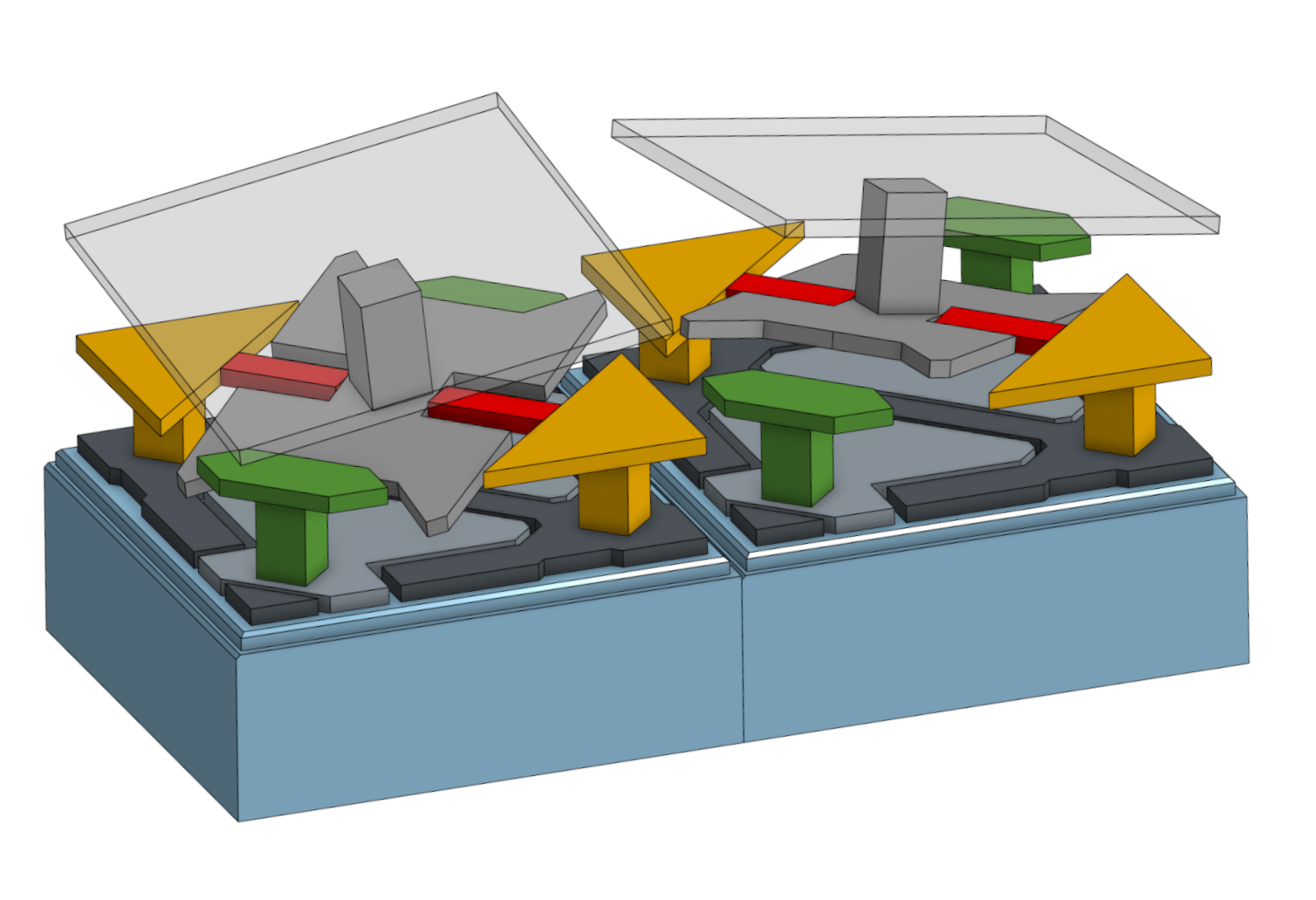}
    \label{fig:pix_right}
  \end{subfigure}
  \caption{
    \textbf{Principle of operation of a DMD in a digital projector.}
    Left, incident light can be reflected towards the projection lens (state {\em on}),
    or onto a beam dump (state {\em off}).
    Right, zoom on the pixels.
    Image adapted from \cite{JacksonDMD}.
  }
  \label{fig:combined_pixel}
\end{figure}

To comprehend both the capabilities and limitations
of DMD technology for coherent wavefront shaping,
it is crucial to understand the device's operating principles
and its original design intentions.
Investigated and developed by Texas Instruments since the 1980s,
DMDs gained prominence in the 1990s for video projection applications since the 1990s
under the commercial name of Digital Light Processing (DLP)~\cite{hornbeck1997digital,Dudley2003emerging}.
The technology enables high-resolution, high-speed, and high-contrast-ratio modulation of light.
DMDs operate by toggling the state of small mirrors between two distinct angles, denoted as $\pm \theta_\text{DMD}$.
The device is originally engineered for amplitude modulation in video projection applications.
In this configuration, one mirror angle directs light into the projection lens,
while the alternate angle results in the light path being blocked (see Fig.~\ref{fig:combined_pixel}).
Given that projectors utilize incoherent light and that the DMD plane is optically conjugated with the projection screen,
aberrations within the DMD plane are generally not problematic.
Similarly, phase fluctuations induced by temperature variations,
as well as minor vibrations from the cooling hardware, are inconsequential in this context.
The DMD is designed to produce binary on/off modulation,
which is then leveraged to generate grayscale images via pulse-width modulation.
Color modulation is accomplished through the use of a color wheel in conjunction with a bright white light source.\\

Third-party companies have developed kits tailored for research applications,
which include a DMD, a control board, and a software interface.
Specifically, Vialux devices~\cite{vialux} offer an FPGA board that enables high-speed modulation
by allowing frames to be stored in the device's memory~\cite{hofling2004alp}.
However, standard Texas Instruments video projector evaluation modules can also be repurposed into wavefront shaping devices~\cite{Cox2021converting},
though at a compromised modulation speed.
These systems can further be converted into phase or complex field modulators.

While other articles exist describing the various aspects of DMDs~\cite{Park2015properties, Scholes2019structured, Cox2021converting, Wang2023diffraction},
this tutorial aims to provide a guide for easily setting up a DMD for wavefront shaping applications in complex media.
In particular, we provide characterization and validation procedures that require minimal changes compared to typical wavefront shaping setups. 
More specifically, we place ourselves in a standard experiment where the goal is to shape the complex wavefront impinging on a complex medium and control or measure its output response. 
This is typically the case for a focusing experiment~\cite{Vellekoop2007focusing} or for measuring the transmission matrix of a complex medium~\cite{Popoff2010Measuring}. 
\red{
In such works,
complex modulation is usually achieved by encoding the optical phase
into the spatial displacement of binary fringes
displayed on the DMD,
followed by filtering high spatial frequencies in the Fourier plane~\cite{lee1979binary}.
Such a configuration permits multi-level complex modulation,
at the cost of a reduced spatial resolution.
The implementation and performance of such systems are discussed in detail in a separate tutorial~\cite{Gutierrez2024DMD}
and are not elaborated further here.
For the remainder of this paper, it will be assumed that the DMD is used
for complex modulation via such a method.
}
\\

We first introduce the diffraction properties of a DMD
and elaborate on how these could impact the system's efficiency.
We also furnish a straightforward criterion for selecting the appropriate DMD parameters for a specified excitation wavelength.
In the next section,
we delve into the aberration impacts brought about by the non-flatness of the DMD surface.
We demonstrate a simple process to characterize this effect and provide compensation solutions.
In the third segment,
we detail the influence of mechanical vibrations that are induced by the DMD's cooling system.
Lastly, we discuss how the thermalization of the DMD chip can potentially result in variations to the DMD response over time.

\section{Choosing the right DMD: Diffraction effects}

A significant distinction between liquid crystal modulators and DMDs
lies in the geometry of the pixel surface: 
The pixels of the DMD are not flat, but are mirrors that can be tilted 
in two different positions around a rotation axis
along the diagonal of the square shape of the pixels.
This difference gives rise to diffraction effects that can adversely affect
both modulation quality and diffraction efficiency, 
\red{defined by the fraction of the incident optical power that 
is redistributed into a single diffraction order.}
The impact of these diffraction effects is highly dependent on several factors:
the wavelength of illumination, the pixel pitch, and both the incident and outgoing angles.
Therefore, in conjunction with selecting an appropriate anti-reflection coating,
it is crucial to ensure that the pixel pitch is compatible with the specific configuration being used.
Texas Instruments offers chips with a variety of pixel pitches $d$,
ranging approximately from $5$ to $\sim25$ \textmu m~\cite{TI}.\\

\subsection{A 1D model}

\begin{figure}
  \centering
  \includegraphics[width = \textwidth]{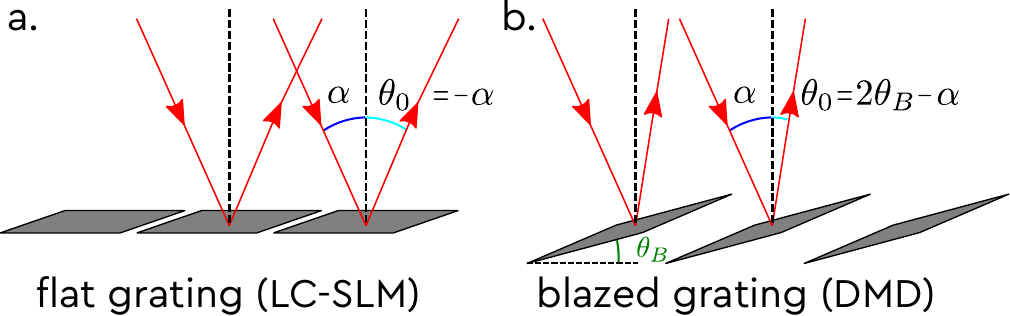}
  \caption{
    \textbf{1D grating geometry.}
    Schematic representation of the geometry of two types of modulators:
    (a) the liquid crystal modulator, equivalent to a flat grating,
    and (b) the DMD geometry, equivalent to a blazed grating.
    $\alpha$ denotes the incident angle relative to the normal of the array plane,
    $\theta_0$ refers to the angle of the zeroth diffraction order,
    and $\theta_B$ is the tilt angle of the mirrors.\\
  }
  \label{fig:grating_geom}
\end{figure}

To achieve a qualitative understanding of this issue,
we consider a 1D array of pixels as illustrated in Fig.~\ref{fig:grating_geom}.
Initially, let's assume that all pixels are in the same state and are illuminated by a plane wave originating from the far field.
Under these conditions, the pixelated modulator essentially functions as a grating,
with a period $d$ that is equivalent to the pixel pitch.
It is important to underscore that these modulators possess a hardware-limited filling fraction,
typically around 90\%.
This translates to an effective active pixel size of $d' < d$.\\
\red{
Here we compare two types of gratings:
Flat gratings, corresponding to LC-SLMs (Liquid Crystal Spatial Light Modulators),
in which all pixels are in the same state;
and blazed gratings, corresponding to DMDs,
in which all pixels are tilted by the same angle.
A grating gives rise to multiple diffraction orders,
whose amplitudes are modulated by a global envelope.
Importantly, these two effects can be decoupled.
The diffraction order angles $\theta_p$
are determined by the grating periodicity,
while the envelope, and in particular the position of its maximum $\theta_{\mathrm{\max}}^{envelope}$,
is governed by the structure of a single pixel,
which constitutes the unit cell of the grating.
}

\subsubsection{Diffraction orders\\}

In general, a grating gives rise to various diffraction orders
with different intensities
and angles $\theta_p$, as dictated by the grating equation:
\begin{equation}
    \text{sin}(\theta_p)+ \text{sin}(\alpha) = p\lambda/d = p \, \text{sin}(\theta_D),    
    \label{eq:1D_orders}
\end{equation}

where $\lambda$ is the wavelength of the light, $\alpha$ is the incident angle, 
$\theta_D = \arcsin \left(\lambda/d\right)$ is the angle of the first diffraction order,
and $p$ is an integer value denoting the orders of diffraction.
The intensity of the individual diffraction orders is influenced by
the response of a single pixel,
constituting the unit cell of the grating,
and that is governed by $d$, $d'$, and its rotation angle in the case of a DMD.
\red{
While this regime is not always relevant,
we can adopt the small-angle approximation to obtain
a qualitative picture of the effect of the incident angle $\alpha$
on the diffraction orders.
In this case, we have $\theta_p \approx p\lambda/d - \alpha$.
Thus, changing the incident angle effectively
rotates all diffraction orders by an angle $-\alpha$.
}\\

\subsubsection{Position of the maximum of the envelope\\}
\red{
The envelope of the diffracted field,
which sets the amplitude of each diffraction order,
is governed by the response of each unit cell, i.e. each pixel.
In a flat grating excited at normal incidence,
the envelope is peaked around the normal to the surface, i.e. $\theta_{max}^{envelope} = 0$.
In a blazed grating,
the tilt angle $\theta_B$ introduces a linear phase ramp
in the response of each individual pixel.
A flat grating can be seen as a particular case with $\theta_B = 0$.
The incidence angle $\alpha$ also adds
a global phase slope.
When the filling fraction approaches 100\%,
i.e. when $d \approx d'$,
the cumulative effect leads to a shift of the envelope
by an angle $2\theta_B - \alpha$.
This results in a maximum of the envelope at
}

\begin{equation}
    \theta_{max}^{envelope}= 2\theta_B - \alpha
    \label{eq:1D_envelope}
\end{equation}

\red{
which reduces to $\theta_{max}^{envelope} = -\alpha$ in the case of a flat grating.
We detail the calculation of this effect in Appendix~A.
}

\subsubsection{Blazed grating condition\\}

\red{
In the case of a flat grating ($\theta_B = 0$), we see from Eqs.~\ref{eq:1D_orders} and~\ref{eq:1D_envelope}
that the position of the first order is always aligned with the maximum of the envelope
at the angle $-\alpha$,
when the filling fraction is close to 1.
In the general case $\theta_B \neq 0$,
$\theta_{max}^{envelope}$ does not, in general, coincide
with any diffraction order~\cite{Park2015properties, Wang2023diffraction}.
This leads to a redistribution of the energy of the incoming beam
into multiple diffraction orders,
resulting in a reduced diffraction efficiency
as well as cross-talk between different pixel states,
as studied in section~\ref{sub:crosstalk}.\\
}

\red{
The maximal diffraction efficiency is obtained 
when the maximum of the envelope matches one order of diffraction. 
This state is achieved when the conditions of the blazed grating equation are fulfilled~\cite{Casini2014on}, 
i.e. when there exist an integer $p$ that satisfies:
}

\begin{equation}
  \text{sin}(2\theta_B-\alpha) + \text{sin}(\alpha)
  = 2 \,\text{sin}(\theta_B)  \,\text{cos}(\theta_B-\alpha)
  = p\frac{\lambda}{d} \, .
  \label{eq:blazed_eq}
\end{equation}

\red{
In particular, in the Littrow configuration, 
i.e. for $\alpha = \theta_B$, we have $\theta_{max}^{envelope} = \alpha$; 
the diffraction and incidence angles are identical.
The optimal condition is then satisfied for $\sin(\theta_B)={p\lambda}/{(2d)}$ $= p\sin(\theta_D)/2$.
}



\subsubsection{Example\\}

\begin{figure}[ht]
  \centering
  \includegraphics[width = 0.75 \textwidth]{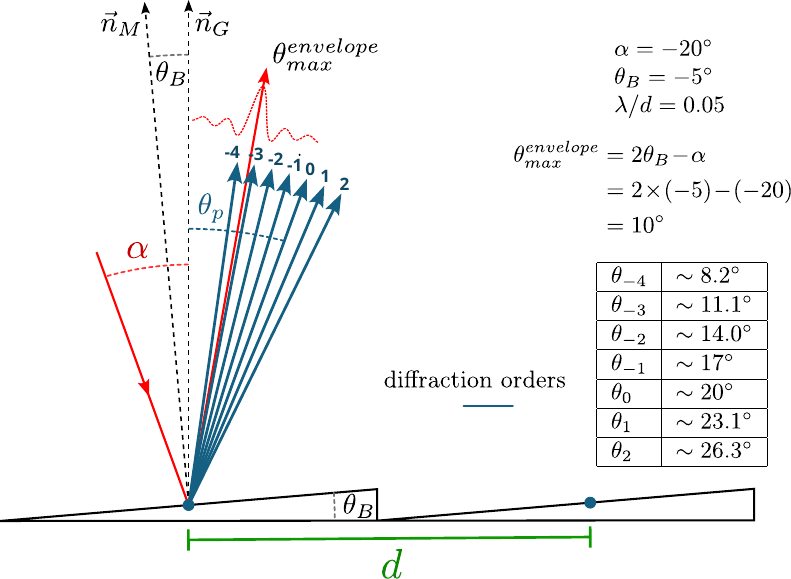}
  \caption{
    \textbf{1D grating example.}
    All angles are defined with respect to the normal of the surface of the the grating, 
    denoted by $\vec{n}_G$, and are taken as positive for clockwise rotations and negative for counter-clockwise rotations.
    The blazing angle $\theta_B$ is the angle between the normal to the micro-mirror $\vec{n}_M$ and $\vec{n}_G$.
    The angles of the diffraction orders $\theta_p$ are independent of $\theta_B$, while the maximum of the envelope is at $\theta_{max}^{envelope} = 2\theta_B -\alpha$.
  }
  \label{fig:grating_geom_example}
\end{figure}

\begin{figure}
  \centering
  \begin{subfigure}{0.49\textwidth}
    \centering
    \includegraphics[width = \textwidth]{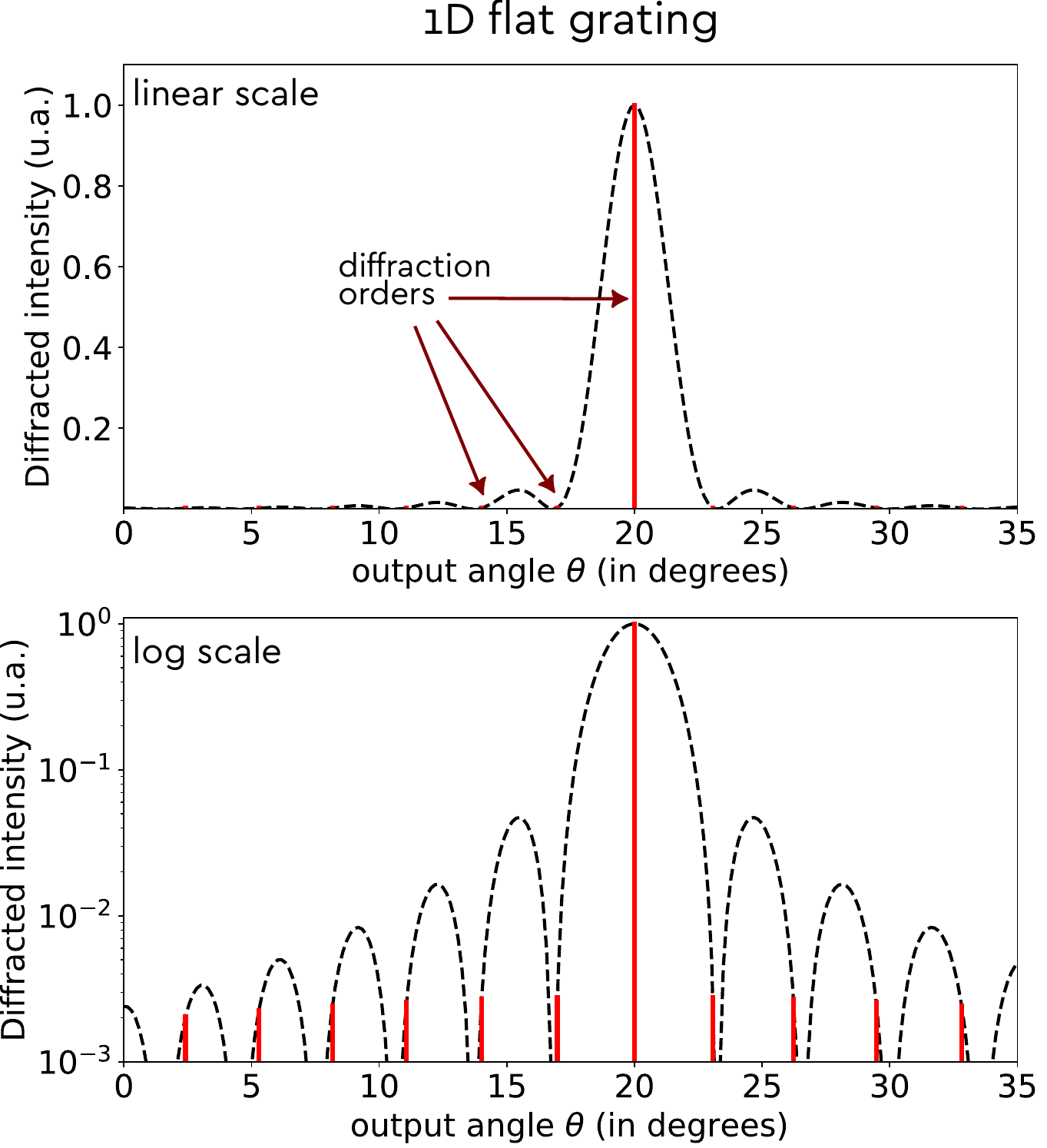}
    \label{fig:flat_grating}
  \end{subfigure}
  \begin{subfigure}{0.49\textwidth}
    \centering
    \includegraphics[width = \textwidth]{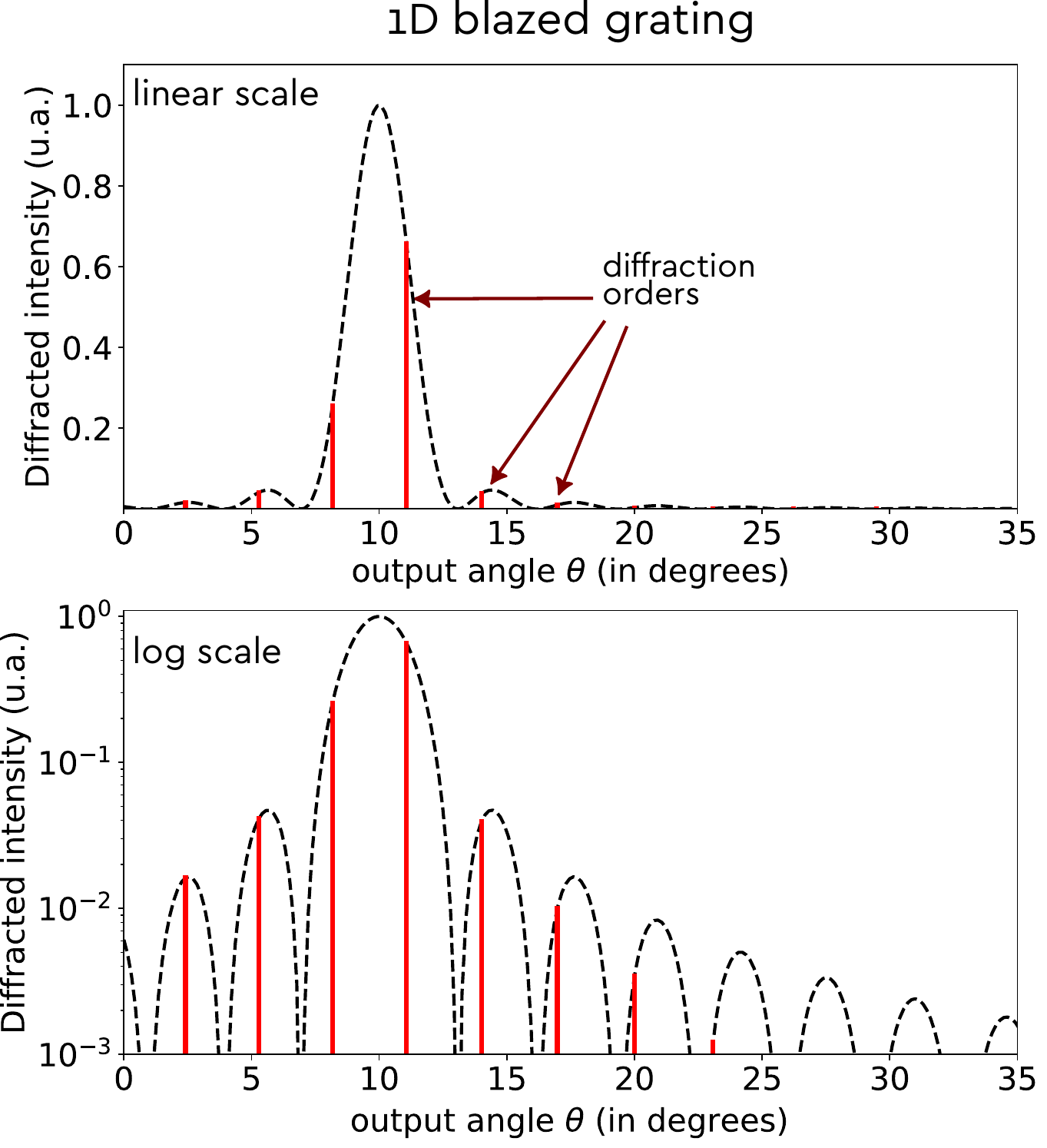}
    \label{fig:blazed_right}
  \end{subfigure}
  \caption{
    \textbf{Flat grating vs blazed grating.}
    Far field diffraction patterns for a 1D flat grating (left) and a 1D blazed grating (right)
    for an input angle of $\alpha = -20^\circ$,
    a filling fraction of 95\% (corresponding to a 2D filling fraction of $\approx 90$\%),
    a pixel tilt angle of $\theta_B = 5^\circ$,
    and a wavelength to pixel pitch ratio $\lambda/d=0.05$.
    Vertical lines represent the angles of the diffraction orders, peaked at $\theta_p$ which satisfy $\sin(\theta_p)+\sin(\alpha)=p\lambda/d$, 
    and the black dashed curve represents the amplitude of the field which peaks at $\theta_{max}^{envelope}=2\theta_B-\alpha$.
  }
  \label{fig:gratings}
\end{figure}

\red{
We represent in Fig.~\ref{fig:grating_geom_example} an example of a geometry 
for a 1D blazed grating and 
in Fig.~\ref{fig:gratings} its angular response 
as well as the one corresponding to a flat grating with the same parameters.
We take a 1D filling fraction of 95\% (correponding to a 2D filling fraction of $\approx 90$\%).
For a flat grating, the zero-th order contains most of the intensity,
the other orders being negligible in comparison.
For the blazed grating example shown,
we are in a situation close to the worst case scenario:
Two diffraction orders have a significant and comparable intensity,
and other orders also have non-negligible contributions.
In the optimal scenario, where the peak of the envelope corresponds to a diffraction order,
it results in a single diffraction order carrying the majority of the energy.
This state is achieved when the conditions of the blazed grating equation are fulfilled~\cite{Casini2014on}:
}



\subsection{The 2D case}

\subsubsection{Blazed grating condition\\}
\red{
A DMD consists of a 2D array of micro-mirrors.
In most DMDs, the rotation axis of the mirrors
is aligned with the pixel diagonals.
For convenience in the alignment and manipulation of the optical setup,
it is preferable to work with incident and outgoing beams
whose optical axes lie in the horizontal plane.
A straightforward and common solution is to rotate the chip by $45^\circ$
with respect to the horizontal plane,
which makes the pixels' rotation axis vertical.
This configuration is depicted in Fig.~\ref{fig:2d_geom}.
We show in Appendix B. that it leads to a new blazed grating equation that reads:
}\\

\begin{figure}
  \centering
  \includegraphics[width = 0.9\textwidth]{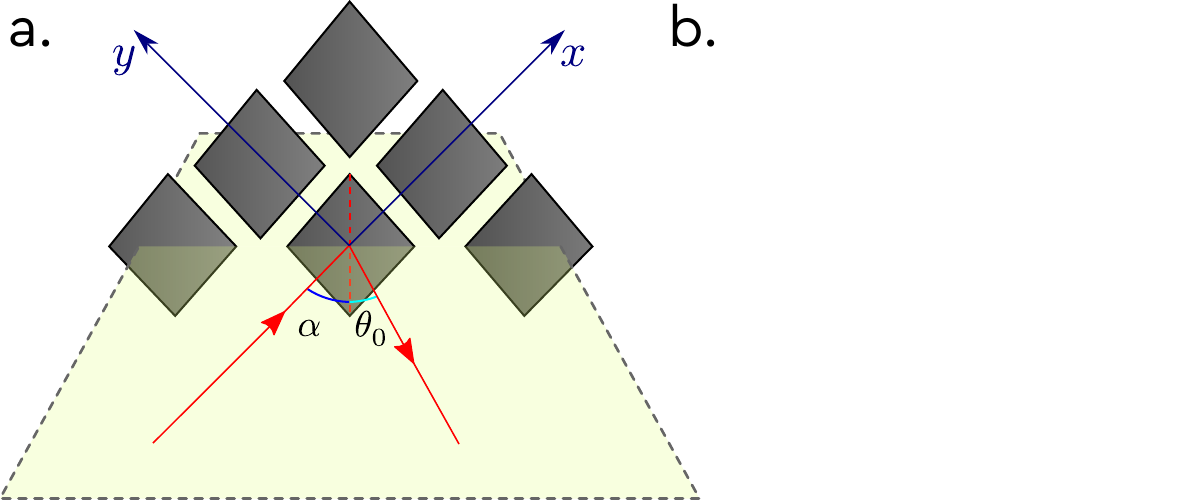}
  \caption{
    \textbf{2D grating geometry.}
    (a) Schematic representation of the geometry of the DMD.
    The incident  angle $\alpha$ and the reflection angle $\theta_{max}^{envelope}$ (defined by the peak of the diffraction envelope) are situated
    within the horizontally plane, illustrated in yellow.
    (b) Photograph of the DMD chip oriented so that the rotation axis of the pixels
    is aligned vertically.
  }
  \label{fig:2d_geom}
\end{figure}

\begin{equation}
  \text{sin}(2\theta_B-\alpha) + \text{sin}(\alpha)
  = 2 \,\text{sin}(\theta_B)  \,\text{cos}(\theta_B-\alpha)
  = p\frac{\sqrt{2}\lambda}{d} \, ,
\end{equation}

with $p$ an integer.
Note that this is what one would obtain using Eq.~\ref{eq:blazed_eq} for 
blazed grating with a pitch $d/\sqrt{2}$, or a 1D grating rotated by 45 degrees.
Those systems are not equivalent in the general case, 
but the grating condition in the 2D case is
similar when considering an incident plane wave in the horizontal plane.\\

\subsubsection{Blazed number\\}
We can quantify how close we are to the ideal case,
i.e., when satisfying the blazed equation,
by defining a {\em blazed number} $\mu$ as introduced in~\cite{WFSnet_diffraction}:

\begin{equation}
  \mu =
  \left| 4 \frac{d}{\sqrt{2}\lambda}
  \left[
    \text{sin}(\theta_B)\text{cos}(\theta_B-\alpha)
    \right]
  \mod{2} -1
  \right| \, ,
  \label{eq:blazed_number}
\end{equation}

where $\mod{2}$ represents the modulo 2 operation.
$\mu$ is maximal and equals $1$ when the blazing equation is satisfied,
i.e. when one order of diffraction contains most of the energy,
\red{ 
corresponding to one diffraction order $p$
being close to the maximum of the envelope
($\theta_p \approx \theta^{\mathrm{envelope}}_{\max}$).
$\mu$ is minimal in the worst-case scenario,
i.e. when four diffraction orders have a significant and equal intensity.
Note that, since the maximum of the envelope is no longer aligned
with the zeroth order, corresponding to the specular reflection on the micro-mirrors,
the value of the order $p$ that aligns with the envelope peak
when $\mu \approx 1$ can take any integer value,
and depends on the DMD parameters and on the angle of incidence $\alpha$.
}
\\

\subsubsection{Simulations\\}
\begin{figure}
  \includegraphics[width = \textwidth]{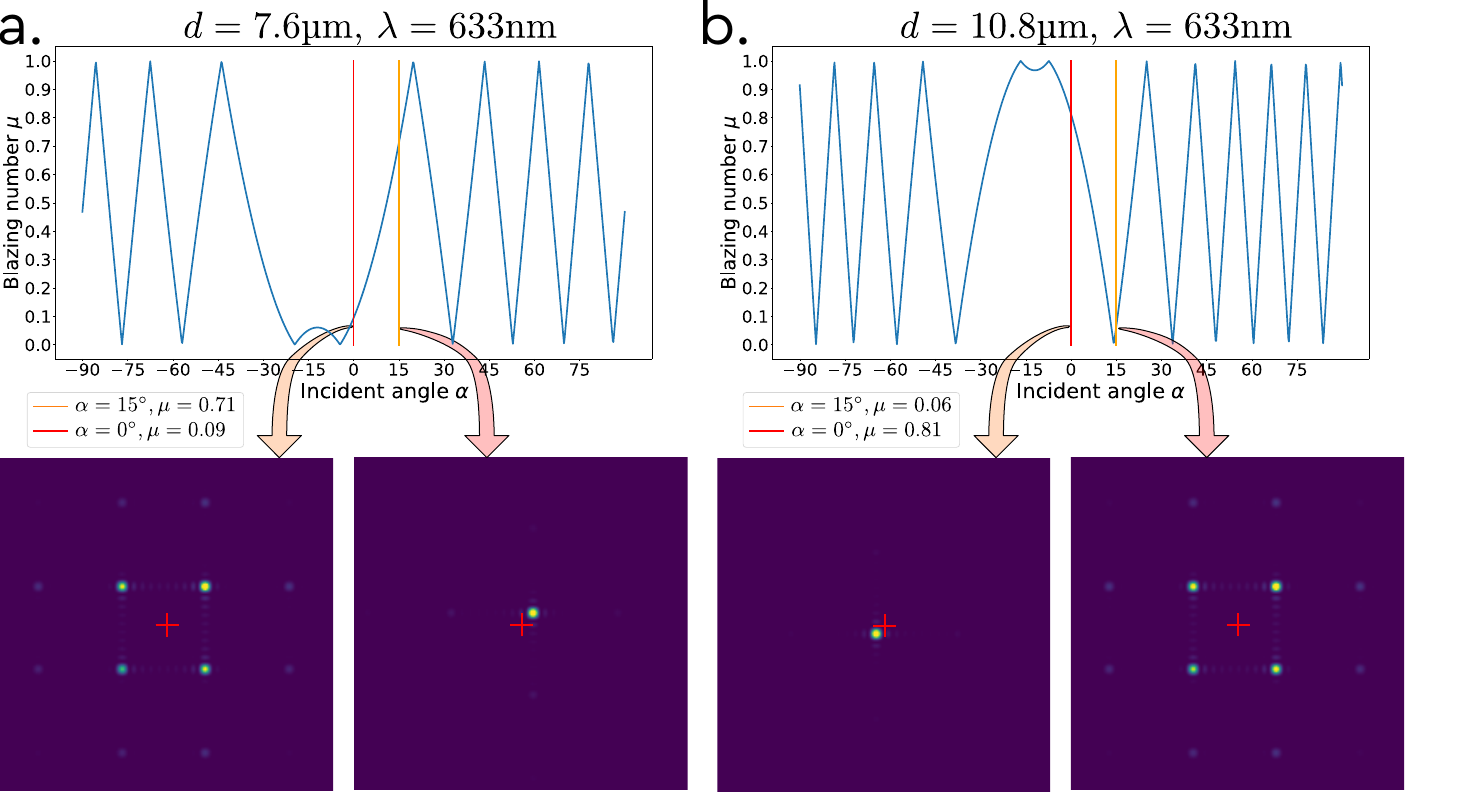}
  \caption{
    \textbf{Blazed number and far-field diffraction patterns.}
    Blazing number $\mu$ (Eq.(\ref{eq:blazed_number})) as a function of the incident angle $\alpha$  (top)
    for a pixel pitch of $d=7.6$\textmu m (a.)
    and $d=10.8$\textmu m (b.).
    Corresponding far-field diffraction patterns (bottom)
    for two incident angles $\alpha = -15^\circ$ and $\alpha = 0^\circ$.
    The red cross indicates the maximum of the envelope $\theta^{envelope}_\text{max} = 2\theta_B - \alpha$
    \red{with $\theta_B = -12^\circ$}.
  }
  \label{fig:mu76}
\end{figure}

\begin{figure}
  \centering
  \includegraphics[width = \textwidth]{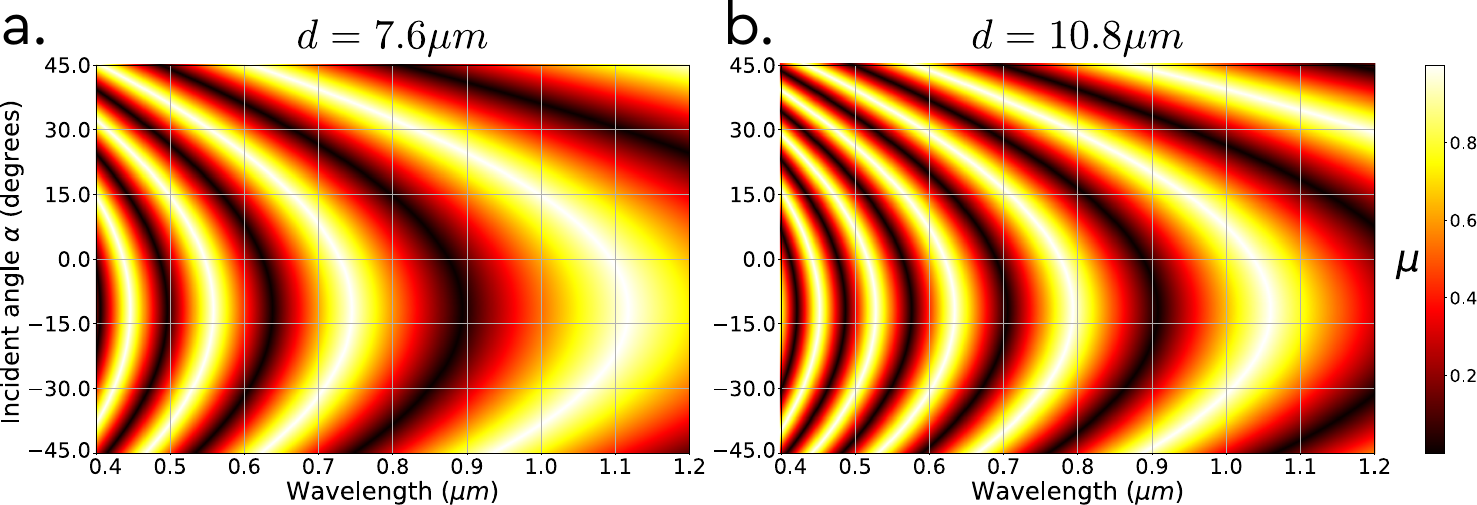}
  \caption{
    \textbf{Blazed number as a function of the wavelength and the incident angle.
    }
    Color map representation of the blazed number for a pixel pitch of $d=7.6$\textmu m (a.)
    and $d=10.8$\textmu m (b.) \red{for $\theta_B = -12^\circ$}.
  }
  \label{fig:mu_2d}
\end{figure}

To demonstrate the effect, we first conduct a simulation of a DMD using Python
(refer to tutorial and code in~\cite{WFSnet_diffraction})
with two pixel pitches of
$d=7.6$\textmu m and $d=10.8$\textmu m,
under a coherent excitation at $\lambda=633$nm.
Fig.~\ref{fig:mu76} shows the estimated blazed number $\mu$
as a function of the angle of incidence $\alpha$ in the horizontal plane,
along with the far field diffraction pattern for two distinct incident angles. 
Fig.~\ref{fig:mu_2d} shows the blazed number as a function of both the incident angle 
and the wavelength for pixel pitches of $d=7.6$\textmu m and $d=10.8$\textmu m.
It should be noted that the efficiency of diffraction,
\red{i.e. the fraction of the incident optical power that is diffracted into a given diffraction order 
relative to the incident power, }
can be altered by adjusting
the angle of incidence $\alpha$.
However, its impact is relatively confined
within an acceptable angular range that aligns with experimental limitations
(i.e. for angles far from $\pm 90^\circ$).
Far-field patterns are centered around the maximum of the envelope
for an output angle $2\theta_B - \alpha$
(marked by a \red{red} cross).
We see that for small positive values of $\alpha$,
the pixel pitch of $d=10.8$\textmu m leads to a blazed number $\mu$
close to $1$.
It corresponds in the far field to having one bright order of diffraction
close to the maximum of the envelope.

\begin{figure}
  \centering
  \includegraphics[width = \textwidth]{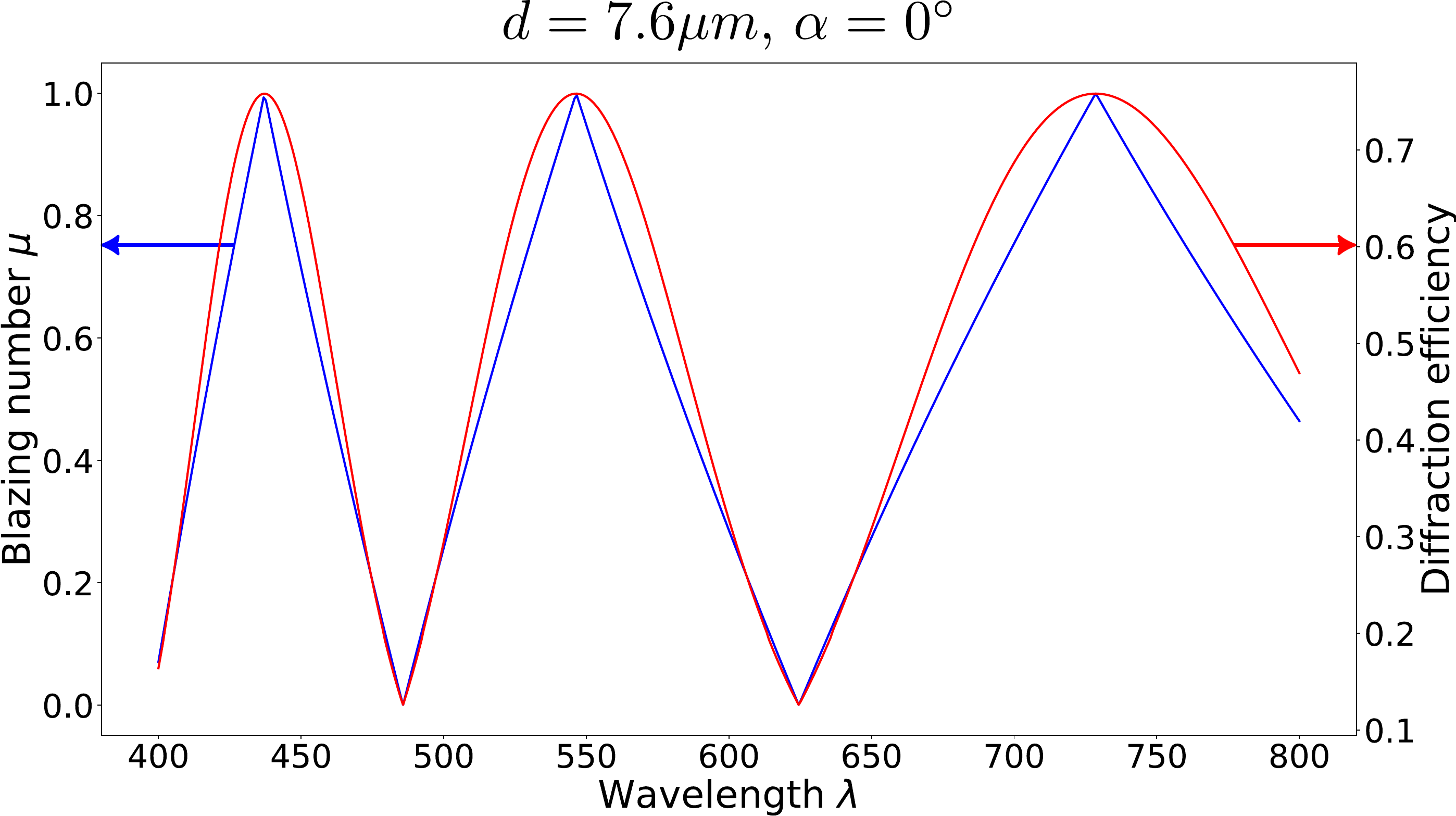}
  \caption{
    \textbf{Blazed number vs diffraction efficiency.
    }
    Plot of the blazed number and the diffraction efficiency for a normal incidence and a pixel pitch of  7.6 \textmu m 
    as a function of the wavelength. 
    The diffraction efficiency is estimated by comparing the energy in a \red{diffraction} limited spot at the 
    location of the brightest diffraction order to the total diffracted energy.
  }
  \label{fig:mu_vs_diff}
\end{figure}

To illustrate the link between the blaze number and the efficiency of the diffraction setup, 
we compute the ratio of the energy in the brightest diffraction spot to the total diffracted energy 
for different values of the wavelength, 
with fixed values for the pixel pitch (7.6 \textmu m) and incident angle (normal incidence).
We simulate the DMD with no gap between the pixels (i.e., a filling fraction of 100\%) 
and with 10 pixels in each direction, 
where all the mirrors of the pixels are in the same state. 
The results are shown in Fig.~\ref{fig:mu_vs_diff}.
We observe that the maxima (resp. minima) of the diffraction efficiency 
correspond to the maxima (resp. minima) of the blaze number. 
Note that the actual values of the minimum and maximum diffraction efficiency depend on the parameters of the DMD, such as pixel pitch, filling fraction, resolution, etc.
\red{
In a practical situation, pixels are not all in the same state:
the configuration is a mix of \textit{on} and \textit{off} states
corresponding to angles $\pm \theta_B$.
The diffraction efficiency then also depends
on the modulation scheme and on the specific parameters used for the modulation.
This topic is addressed in detail in a separate tutorial
focusing on complex modulation using binary amplitude modulators~\cite{Gutierrez2024DMD}.
}\\

\subsection{Experimental measure of the optimal incident angles}

\begin{figure}[!ht]
  \centering
  \includegraphics[width = \textwidth]{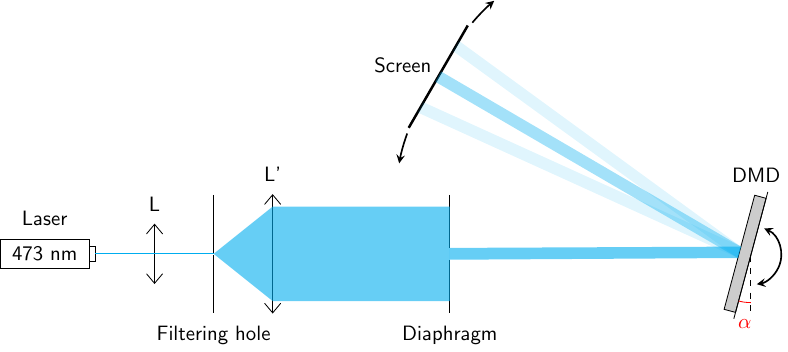}
  \caption{
    \textbf{Blazed number maxima measurement set up.
    }
    A laser beam is filtered and expanded using a telescope (objective L and lens L'). A diaphragm is used to adjust the beam diameter before it reflects off the DMD with all pixels in \textit{on} (resp. \textit{off}) state. A movable screen allows to see the brightest diffraction orders as the DMD is rotated. 
  }
  \label{fig:setup_diff}
\end{figure}

We experimentally characterized a DLP9500 DMD with a pixel pitch of $d = 10.8$\textmu m at $\lambda = 473$nm. The illumination geometry and DMD orientation are as presented in Fig.\ref{fig:2d_geom}. The experimental setup, shown in Fig.\ref{fig:setup_diff}, consists of an objective (L), a lens (L'), and a pinhole to produce a clean, collimated laser beam. A diaphragm is used to control the size of the beam incident on the DMD, and the reflected light is observed on a screen.

We then illuminate  with a collimated monochromatic beam the DMD in its full-\textit{on} (resp. full-\textit{off}) state, 
corresponding to all pixels being rotated by an angle $\theta_B$ (resp. $-\theta_B$), 
with $\theta_B=12$ degrees.
The light distribution across the diffraction orders can be assessed directly with the naked eye by placing a screen after reflection from the DMD. 
The DMD is mounted on a rotating plate with its axis of rotation orthogonal to the table, allowing variation of the angle $\alpha$, as defined in Fig.~\ref{fig:2d_geom}. The reference angle ($\alpha = 0$) is determined by auto-collimating the incident beam with the DMD in its idle state ($\theta_B = 0$).

\begin{figure}
  \centering
  \begin{subfigure}{0.8\textwidth}
    \centering
    \includegraphics[width = \textwidth]{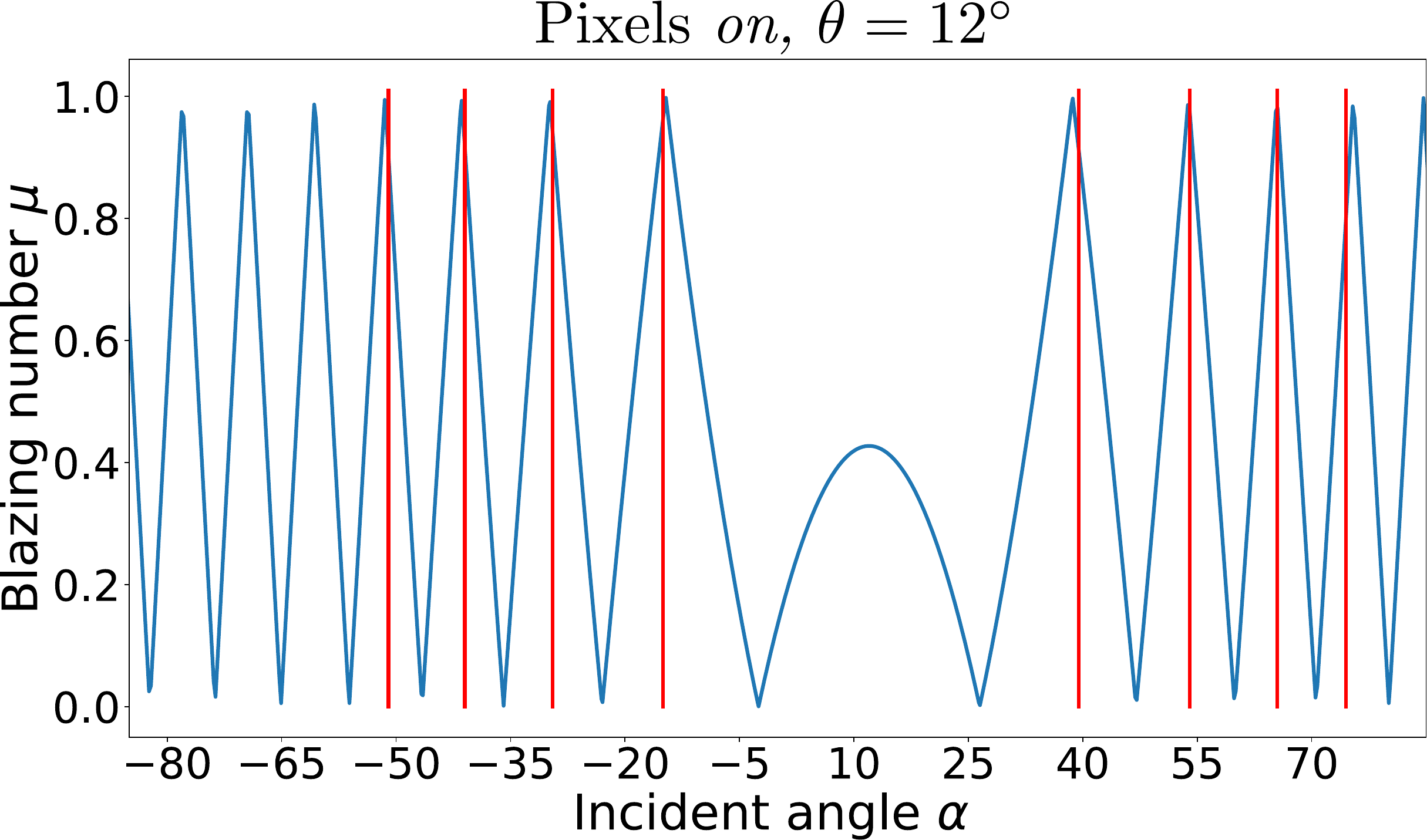}
  \end{subfigure}
  \begin{subfigure}{0.8\textwidth}
    \centering
    \includegraphics[width = \textwidth]{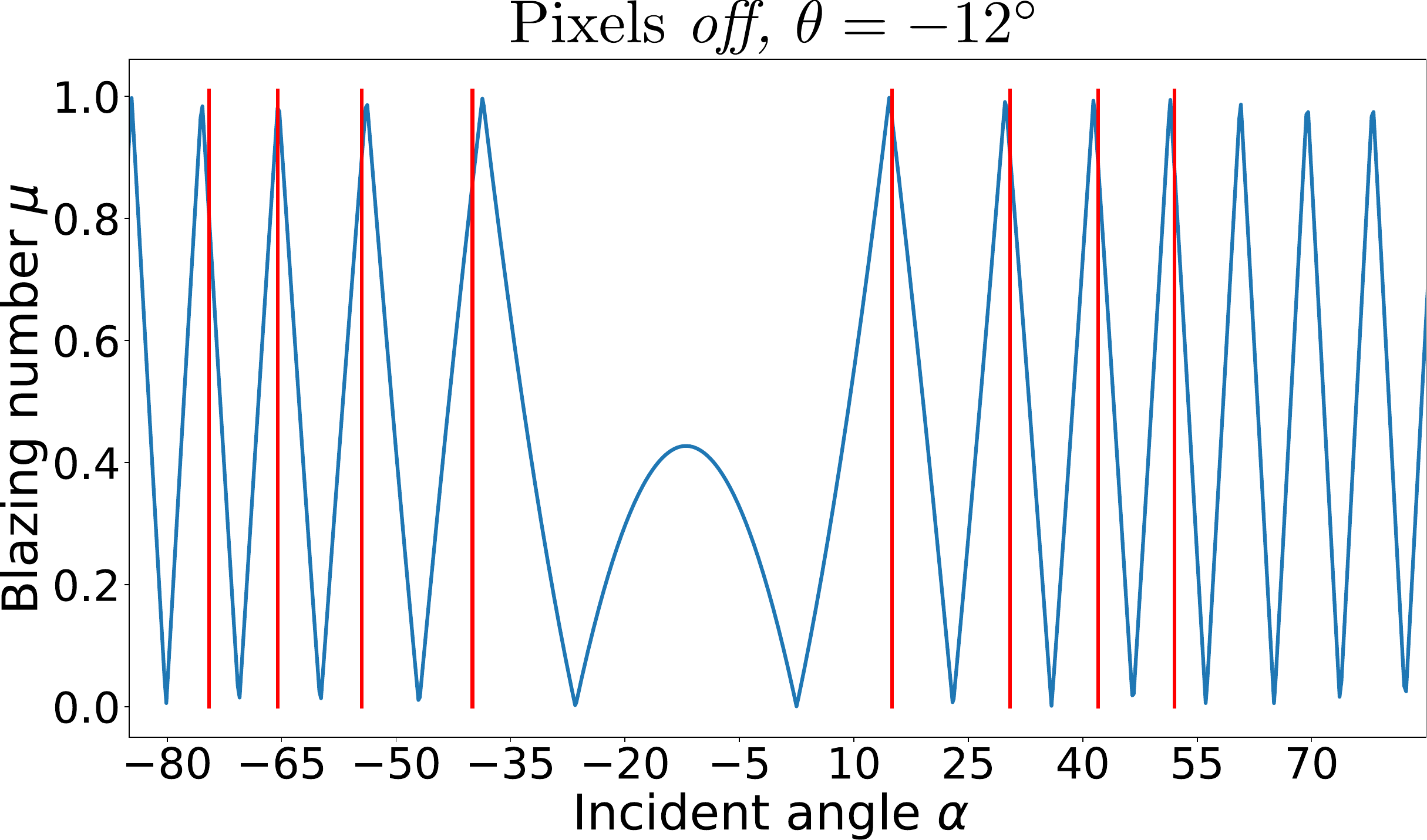}
  \end{subfigure}
  \caption{
    \textbf{Optimal angles for diffraction efficiency, experiment vs predictions.}
     Blazing number $\mu$ as a function of the incident angle $\alpha$  (top)
    for a pixel pitch of $d=7.6$\textmu m
    and $\lambda = 473\text{nm}$, 
    when all the pixels are \textit{on}-state (resp. \textit{off}-state), 
    corresponding to a tilt angle $\theta=12^\circ$ (resp. $\theta=-12^\circ$).
    Blue curves represent the predictions of Eq.(\ref{eq:blazed_number}) 
    and red curves the experimentally measured angles for which the 
    energy is maximal in one diffraction spot.
  }
  \label{fig:exp_vs_theo}
\end{figure}

The DMD is then rotated until the reflected light is almost entirely on a single diffraction order, with adjacent orders extinguished. 
The corresponding values of $\alpha$  are shown in Fig.\ref{fig:exp_vs_theo} as red lines, 
illustrating a good correlation between the computed evolution of 
the blazed number $\mu$ (Eq.\ref{eq:blazed_number}) 
and the diffraction efficiency.

\subsection{Modulation cross-talk}
\label{sub:crosstalk}

\begin{figure}
  \centering
  \includegraphics[width = 0.95\textwidth]{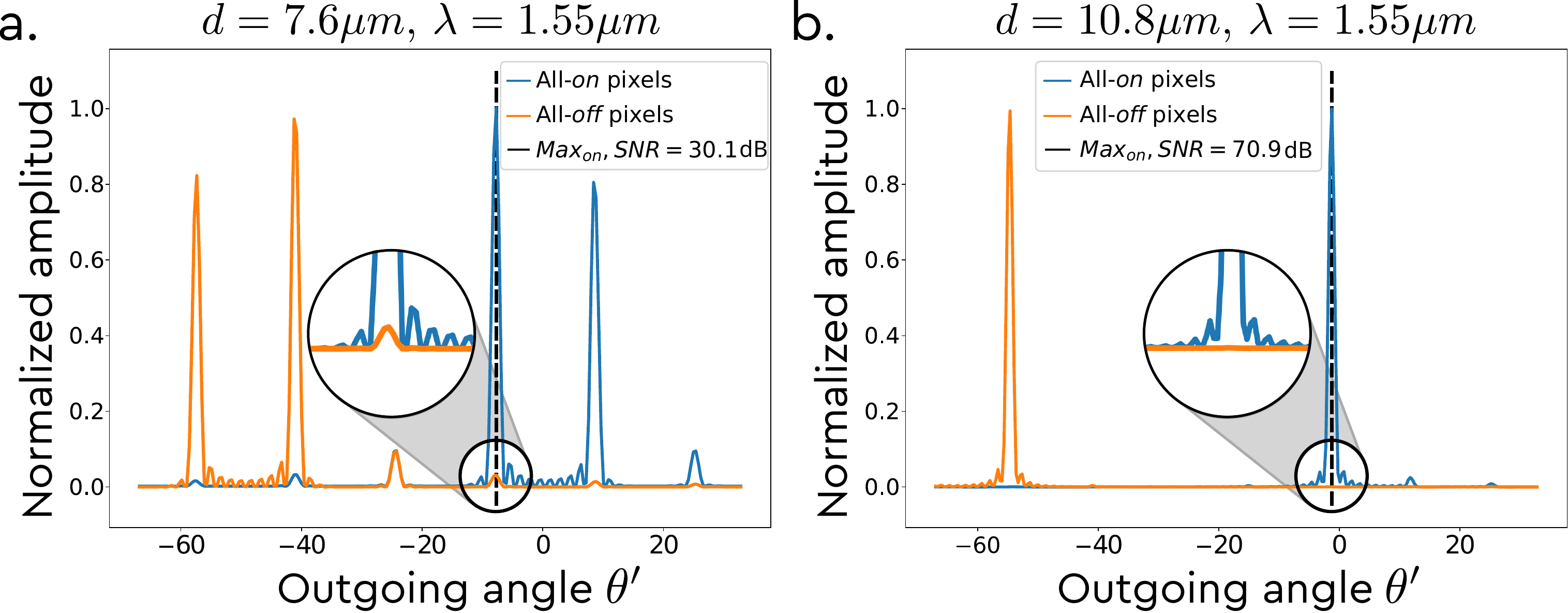}
  \caption{
    \textbf{Cross-talk between {\em on} and {\em off} states.}
    We show the computed normalized amplitude of the diffraction patterns corresponding
    to all the pixels in the {\em on} state (blue)
    and {\em off} state (orange)
    for two different pixel pitches,
    $d = 7.6$\textmu m (a.)
    and $d = 10.8$\textmu m (b.)
    with the same experimental conditions.
    In the first case, $\mu$ is close to $0$,
    we observe a non-negligible contribution of the {\em off} state
    at the main diffraction order of the {\em on} state,
    thus creating unwanted cross-talk.
    In the second case, $\mu$ is close to $1$,
    the contribution of the {\em off} state is negligible.
  }
  \label{fig:xtalk}
\end{figure}

In practice, we place a pinhole or iris to select
one order of diffraction,
corresponding to the {\em on} state.
Having a small value for the blaze number $\mu$
not only restricts the amount of light modulation
due to the diminished diffraction efficiency,
it also influences the modulation quality by inducing cross-talk
between the two states of the DMD pixels.
Until now, we have assumed that all the pixels are in the same state.
In actual usage of the DMD,
it becomes necessary to modulate the state of each pixel individually.
When $\mu$ approaches zero, higher orders of diffraction
still possess a significant intensity,
as demonstrated in the 1D case in Fig~\ref{fig:gratings}.
One adverse implication is that pixels in the {\em off} state
may contain orders of diffraction that are not blocked by the pinhole,
and therefore, will contribute as an interference to the modulated wavefront.
We show in Fig.~\ref{fig:xtalk} the normalized amplitude of the diffraction patterns
corresponding to all the pixels in the {\em on} (blue curve) and {\em off} (orange curve) states
for the same experimental conditions but with two different pixel pitches
leading to situations close to the worst (a) and best (b) case scenarios.
We observe the presence of a non-negligible contribution of the {\em off} state
at the main diffraction order of the {\em on} state in the first case.
While this contribution might appear weak,
it does affect the quality of the modulation
since the modulation scheme typically necessitates about half the pixels
to be in the {\em off} state for phase modulation~\cite{lee1979binary},
and even more so for elaborate modulation schemes~\cite{goorden2014superpixel,Gutierrez2024DMD}.\\

\subsection{Dispersion}

\begin{figure}[ht]
  \centering
  \includegraphics[width = 0.75\textwidth]{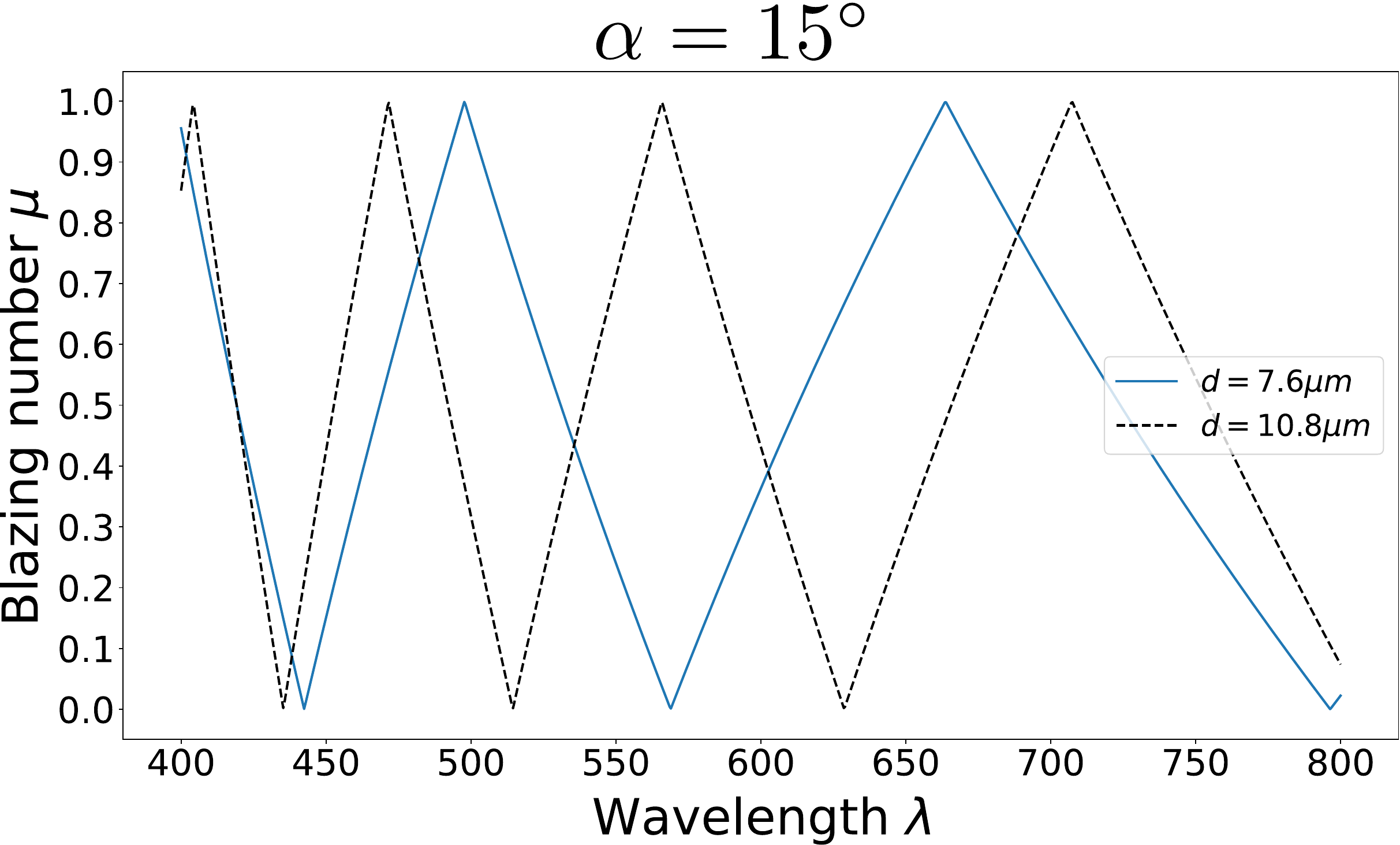}
  \caption{
    \textbf{Dispersion of the diffraction effect.}
    Blazing number $\mu$ (Eq.(\ref{eq:blazed_number})) as a function of the incident wavelength
    for an incident angle $\alpha = 20^\circ$
    and a pixel pitch of $d=7.6$\textmu m (blue curve)
    and $d=10.8$\textmu m (dashed black curve).
  }
  \label{fig:dispersion}
\end{figure}

DMDs are composed of metallic small mirrors,
the response of which is minimally affected by wavelength changes.
This is particularly advantageous for broadband applications requiring amplitude modulation
and operating on a plane conjugated to the DMD's surface.
This is the case for the originally intended application of video projection.
However, for wavefront-shaping applications,
it is typically required to select a specific diffraction order
to acheive phase or complex modulation~\cite{lee1979binary,Gutierrez2024DMD}.
Under such circumstances,
the wavelength-dependency of the diffraction effect becomes important.
The blazed number, denoted by $\mu$ (according to Eq.(\ref{eq:blazed_number})),
scales inversely with the wavelength.
Fig.~\ref{fig:dispersion} shows the blazed number $\mu$
as a function of the wavelength for two pixel pitches $d=7.6$\textmu m and $d=10.8$\textmu m,
with an incident angle of $\alpha = 20^\circ$.
Within the visible spectrum, $\mu$ fluctuates between $0$ to $1$ over a typical range of roughly 100 nm.\\

\begin{tldr}
  \textbf{TL;DR:}\\
  DMDs act as blazed gratings.
  For a given operation wavelength,
  we need to find the right pixel pitch
  to have a good modulation quality
  and diffraction efficiency.
  It can be done by estimating the \textit{blazed number} $\mu$ introduced in eq.~(\ref{eq:blazed_number})
  or directly using our custom online tool~\cite{popoffDMDDiffractionTool}.
\end{tldr}

\subsection{Python code example}

We provide in the paper repository~\cite{github} a Python code
to simulate the diffraction effect of a DMD
by computing the far field pattern for a set of realistic parameters.
It provides a simple way to estimate the blazed number $\mu$ introduced in eq.~(\ref{eq:blazed_number})
to assess the quality of the modulation at the desired wavelength
for a given pixel pitch.
We also propose an online tool accessible at
\url{https://www.wavefrontshaping.net/post/id/49}.

\section{Characterizing aberration effects}

\subsection{Presentation of the problem}

While only capable of providing hardware binary amplitude modulation,
DMDs serve as a potent tool for wavefront shaping and sensing.
These applications  critically require characterization and correction of aberrations
caused by the non-flatness of the DMD surface.
For LC-SLMs, the manufacturer typically characterizes the surface's inhomogeneities within the plane of the modulator and
provides a spatial phase profile of the introduced aberrations based on the operational wavelength.
It is noteworthy that when utilized for intensity modulation on a plane conjugated to the DMD plane,
as it is done in digital projectors,
the system becomes insensitive to the aberrations caused by the DMD surface.
Consequently, these effects are commonly overlooked and rarely documented in the information provided by the manufacturers. 
\red{
However, studies suggest that
the residual stresses developed in the thin films
during the manufacturing process of the DMDs
induce non-negligible surface deformations~\cite{WEI2004375, Cao2001curvature, Brown2021multicolor}.
}

\subsection{Finding the correction pattern}

In the literature, various methods have been proposed to characterize
the phase pattern of the DMD aberrations in the plane of the modulator.
Typically, this involves using a model for the aberrations,
tweaking the parameters to align with the measurements~\cite{Matthes2019Optical,Scholes2019structured, Brown2021multicolor}.
An alternative approach entails direct measurement of the distorted wavefront,
either employing a wavefront sensor such as a Shack-Hartmann~\cite{Lee2023compensation},
or via interferometry.
While such methods yield accurate results,
they necessitate adaptation to the particular
setup conditions,
and frequently require supplementary optical components,
meticulous alignment,
and custom software.
\red{
Moreover, once the phase correction map has been obtained,
it must then be aligned with the pixel array
in order to implement the phase correction on the DMD
and compensate for aberrations.
}
In this section, we introduce a straightforward method for characterizing aberrations using a lens and a camera.
\red{
The goal is to \textit{in situ} compensate for the aberrations
using a phase modulation scheme (Lee holograms),
by iteratively optimizing the point spread function (PSF)
at the focal plane of a lens. 
The resulting phase correction map is the conjugate of the one
induced by the surface deformation of the DMD,
and can subsequently be used to correct aberrations
with any desired complex modulation scheme.
}
This technique can be employed for any system that offers phase modulation,
such as LC-SLMs or deformable mirrors. \\

\begin{figure}
  \centering
  \includegraphics[width = 0.95\textwidth]{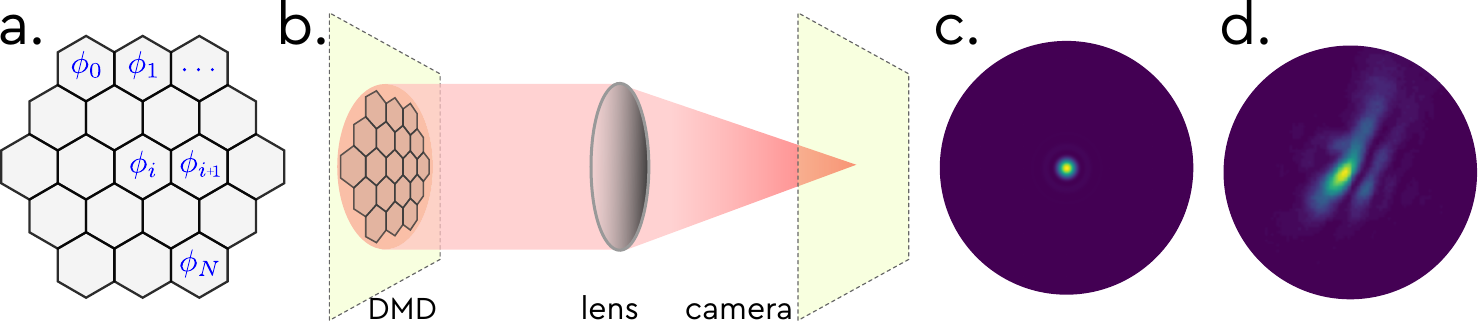}
  \caption{
    \textbf{Setup for aberration correction.}
    (a) A schematic representation of the DMD array divided into macropixels,
    of which the phase can be controlled.
    The precise shape of the macropixels and the area controlled are not crucial for the characterization of aberrations to be effective,
    provided that the spatial sampling is sufficient to capture the highest spatial frequencies of the aberrations.
    (b) A schematic representation of the setup employed to characterize the aberrations.
    A camera is used to image the far field on the DMD surface using a lens.
    (c) shows the ideal intensity pattern corresponding to the numerical aperture of the illumination setup,
    i.e. its theoretical ideal PSF.
    (d) show the actual recorded intensity pattern corresponding to a distorted PSF due to the DMD aberrations.
  }
  \label{fig:dmd_aberr_setup}
\end{figure}

\begin{figure}
  \centering
  \includegraphics[width = 0.95\textwidth]{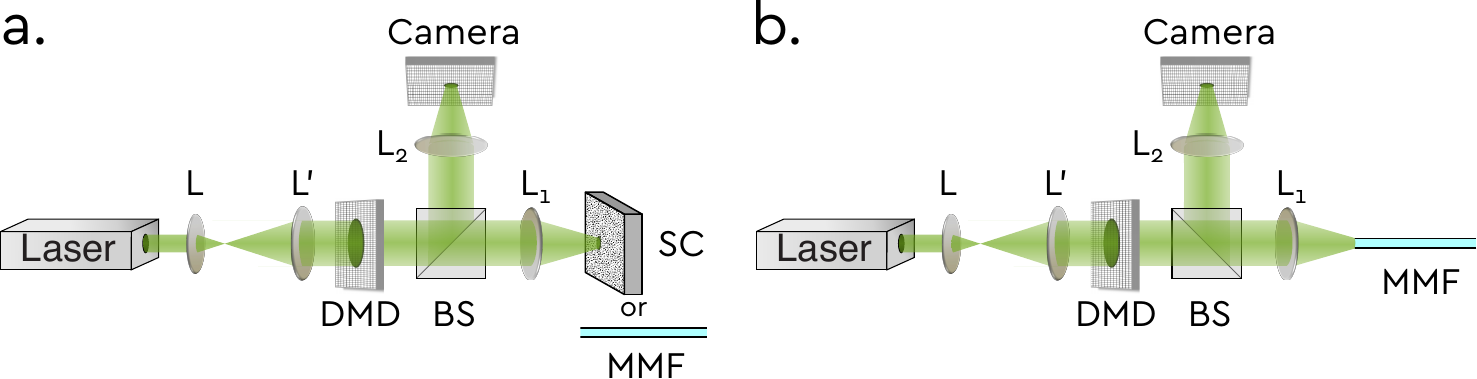}
  \caption{
    \textbf{Detailed aberration characterization setup.}
    (a) A laser beam is expanded and collimated using a telescope (lenses L and L').
    The DMD is represented in transmission for simplicity.
    Light reflected from the DMD is focused by a lens L$_\text{1}$
    onto the complex medium to study, namely a scattering medium (SC)
    or a multimode fiber (MMF).
    A beamsplitter is used to image the far field of the DMD onto a camera
    using a lens (L$_\text{2}$).
    The intensity pattern, up to a \red{homothetic} transformation,
    represent the input excitation on the medium
    which also corresponds to the PSF of the illumination setup.
    (b) In the case of a multimode fiber,
    one can take advantage of the reflection of the input facet
    to directly image the intensity pattern of the input plane.
  }
  \label{fig:dmd_aberr_setup2}
\end{figure}

We assume that the DMD is configured to deliver phase modulation~\cite{lee1979binary,Gutierrez2024DMD}. 
\red{
It is classically done using Lee holograms or similar approaches,
where the optical phase is encoded in the local spatial displacement
of a regular array of amplitude fringes displayed on the DMD.
By filtering a single diffraction order,
the resulting field retains only the phase modulation.
Specific implementations and variants of such approaches are discussed in~\cite{Gutierrez2024DMD}.
}
This implies that the modulator can be divided into $N$ sections,
which we designate as {\em macropixels},
where the phase can be controlled independently.
We use a lens and a camera in its Fourier plane,
illuminating the modulator with a collimated beam
that extends over the entire area of the modulator intended for use.
In scenarios where there are minimal or no aberrations,
the intensity pattern observed would mimic the PSF of the lens,
such as an Airy disk depicted in Fig.~\ref{fig:dmd_aberr_setup}.c.
However, in practice, we encounter a substantially distorted pattern,
like the one represented in Fig.~\ref{fig:dmd_aberr_setup}.d.
A more detailed depiction of the setup for aberration characterization,
in the context of a wavefront shaping application in complex media is presented in Fig.~\ref{fig:dmd_aberr_setup2}.a.
In the case where the medium is a multimode fiber,
one could leverage the reflection from the input surface
to directly visualize the intensity pattern of the input plane,
as shown in Fig.~\ref{fig:dmd_aberr_setup2}.b. 
\red{
Note that for the later approach to be reliable, it is necessary to properly align the system to ensure that the plane of the input facet corresponds exactly to the Fourier plane of the DMD, and that the input facet plane is conjugated with the camera plane.
}
\\

We hypothesize that the aberrations brought about by the DMD are smooth,
and can be depicted by a phase pattern $\phi^\text{aber}$ in the plane of the DMD array.
This could be feasibly approximated by a finite number of Zernike polynomials
$Z_n(r,\theta)$~\cite{Zernike1934beugungstheorie}, as follows:

\begin{equation}
  \phi^\text{aberr}(r,\theta) \approx \sum_{n=0}^N a_n Z_n(r,\theta) \, .
\end{equation}

The goal is to find and display the phase value
$\phi_i^\text{corr}$ on each macropixel $i$ that best compensates for the aberrations,
i.e. $\phi_i^\text{corr} = -\phi^\text{aber}(r_i,\theta_i)$.
We create this pattern in the basis of Zernike polynomials

\begin{equation}
  \phi_i^\text{corr} = \sum_{n=0}^N a'_n Z_n(r,\theta) \, ,
\end{equation}

\noindent the best correction is obtained for

\begin{equation}
  a'_n = -a_n \,\, \forall n \in [0..N]\, .
\end{equation}

\begin{figure}
  \centering
  \includegraphics[width = 0.95\textwidth]{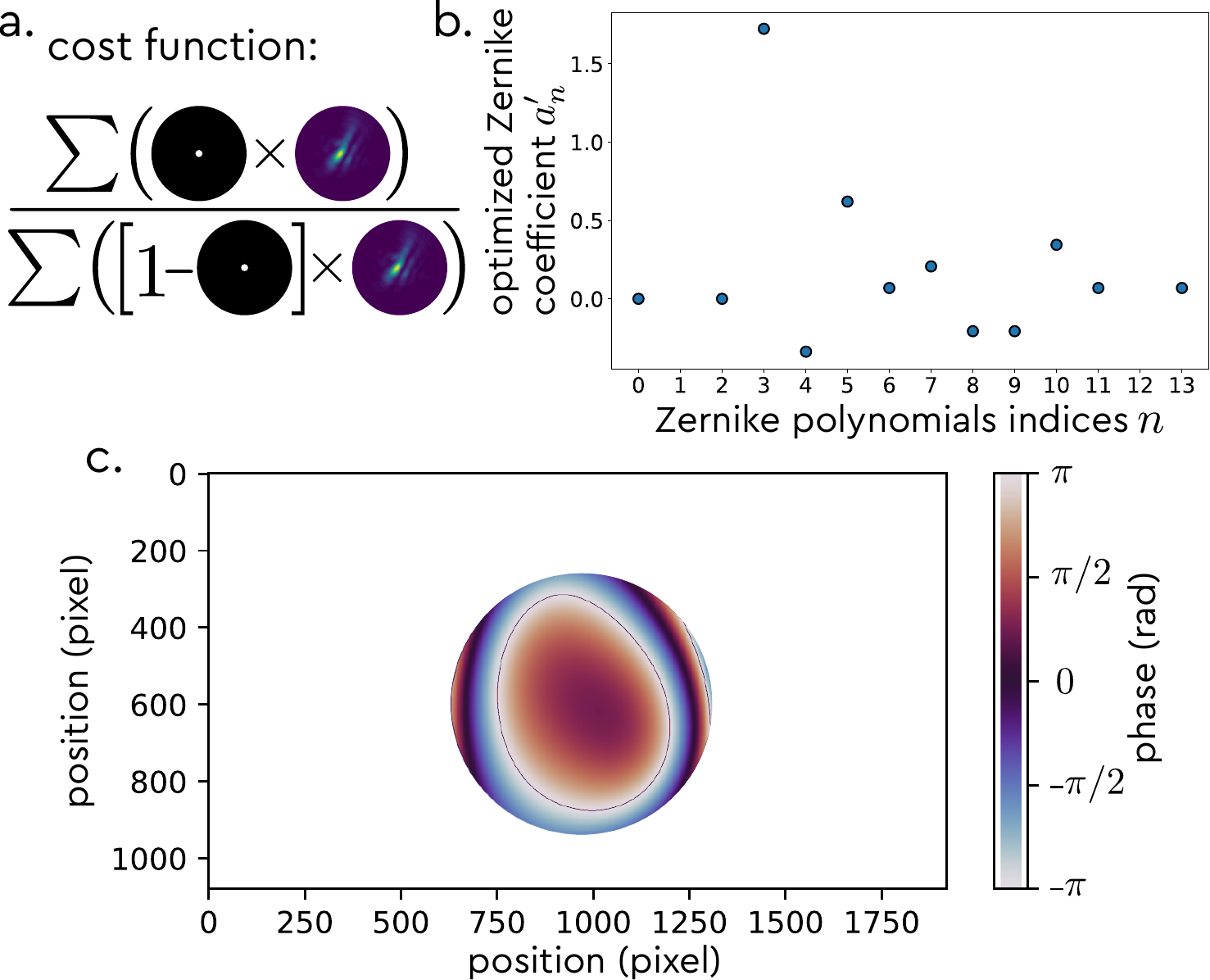}
  \caption{
    \textbf{Correction phase map.}
    (a) Schematic representation of the function we want to optimize, 
    i.e. we optimize the ratio between 
    the sum of the intensity of the measured image in a diffraction-limited disk
    and the sum of the intensity elsewhere.    
    We both want to maximize the intensity of the main lobe of the PSF
    and minimize its side lobes.
    (b) Values of the experimentally obtained optimal coefficients in the Zernike polynomials basis.
    (c) Resulting optimal phase pattern in the chosen illumination area.
  }
  \label{fig:phase_corr}
\end{figure}

We perform a sequential optimization of parameters $a_n$
to maximize a specific function designed
to be maximal for the ideal correction of optical aberrations.
\red{
The approach is inspired by methods used for focusing through
complex disordered media}~\cite{Vellekoop2007focusing}\red{,
by effectively treating the aberrations of the DMD
as equivalent to propagation through an unknown linear medium.
For each degree of freedom
(i.e. the phase on one group of pixels in Ref.~\cite{Vellekoop2007focusing},
or the coefficient of one Zernike polynomial in this section),
we scan different values and retain the one that
maximizes the intensity at the target location.
}
We first  generate a mask, represented by a disk,
centered around the point of maximum intensity of the original image (refer to Fig.~\ref{fig:dmd_aberr_setup}.d)
with a radius equivalent to a single speckle grain.
The exact size of this radius for a successful optimization is not critical
and can be determined by approximating the dimensions of the ideal PSF,
expressed as $r_0 \approx M \frac{\lambda}{2 NA}$,
where $NA$ is the numerical aperture of the optical system and $M$ refers to its magnification.
For a given output intensity pattern, we compute an element-wise product between this image
and the created mask, followed by a summation.
This calculated sum is then divided by an analogous product, but with the complementary mask substituted in place of the original
in order to minimize the side lobes of the PSF.
We set initial parameter values as $a'_n = 0\,\, \forall n \in [0..N]$.
For each parameter, we test different values of $a'_n$,
construct the phases for every micropixel according to
$\phi_i^\text{corr} = \sum_{n=0}^N a'_n Z_n(r_i,\theta_i)$,
record the resulting intensity profile,
and evaluate the corresponding cost function.
For each parameter, the value that results in the maximum output is retained.
To mitigate potential noise or instability, this complete process is reiterated $3$ times for each parameter.\\

To estimate the quality of the correction,
we compute the Strehl ratio of the PSF.
It is defined as the maximum of the measured PSF divided
by the maximum of the ideal one.
This optimal PSF is
the  squared modulus of the Fourier transform of a circular aperture~\cite{airy1835diffraction}:

\begin{equation}
  PSF_\text{ideal} \propto
  \left[
    J_1\left(k a \frac{R}{\sqrt{R^2+f^2}}\right)
    \right]^2 \, ,
\end{equation}

with $J_1$ the Bessel function of the first kind of order $1$,
$k = 2\pi/\lambda$ the wavenumber,
$a$ the radius of the aperture,
$R$ the radial coordinate in the Fourier plane,
and $f$ the focal length of the lens.\\

\begin{figure}
  \centering
  \includegraphics[width = 0.95\textwidth]{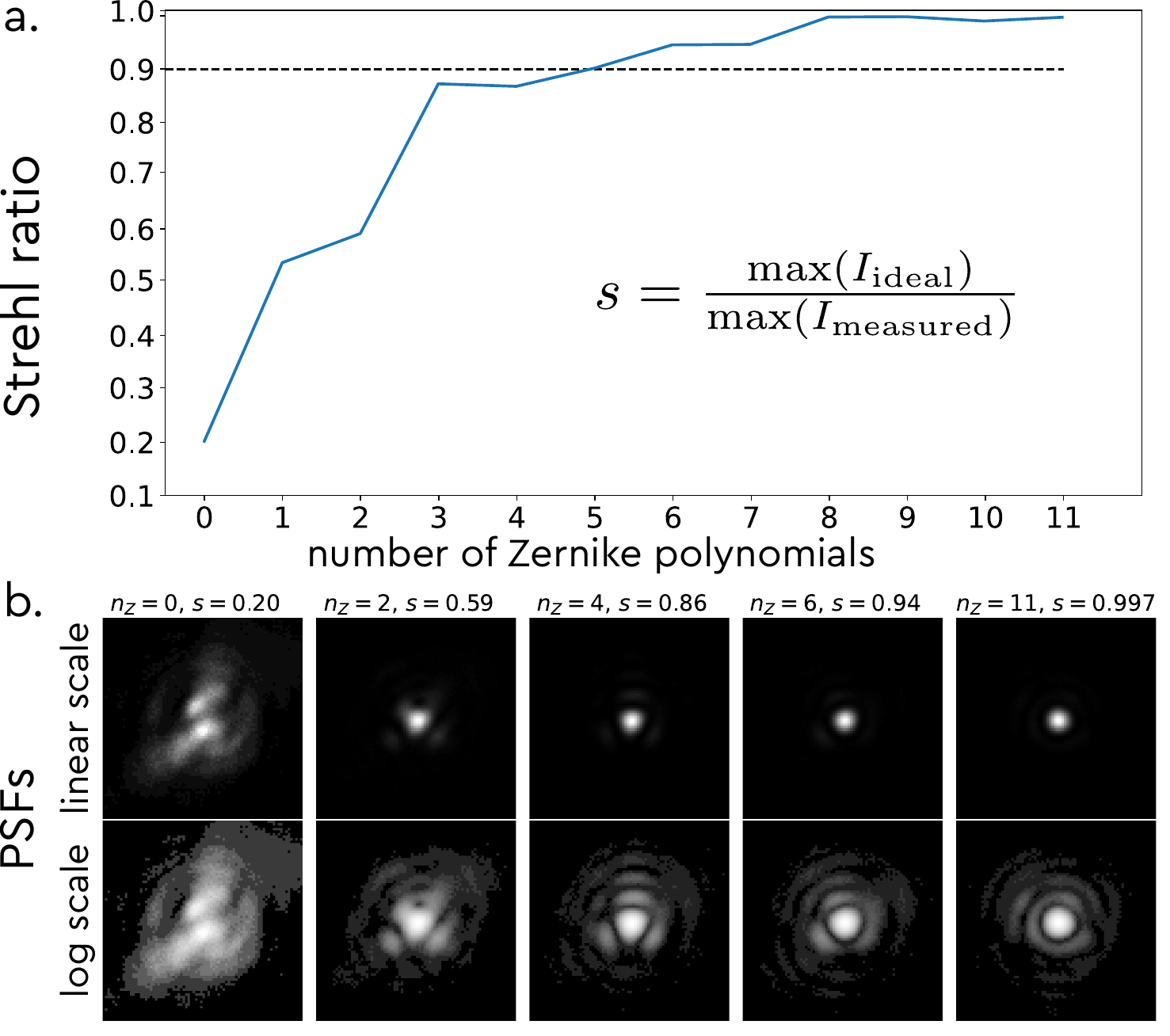}
  \caption{
    \textbf{Effect of Zernike polynomials on the optimization of the PSF.}
    (a) Strehl ratio of the PSF as a function of the number of Zernike polynomials used
    for the compensation of the aberrations.
    (b) Intensity profile of the PSF for different number of Zernike polynomials
    in linear and logarithmic scale.
    To generate this data, we take the final solution of the optimization experimental procedure,
    i.e. the optimal coefficients $a'_n$,
    and generate the corrected PSF with an increasing number of Zernike polynomials.
  }
  \label{fig:zernike}
\end{figure}

As an illustration, we conduct an optimization procedure using 11 Zernike polynomials.
We use a V-9501 Vialux DMD with a DLP9500 TI chip
of resolution 1920 by 1200 pixels and a pixel pitch of 10.8 \textmu m.
The optimization is performed on a disk of radius $340$ pixels,
The illumination is done using an expanded laser beam at 633 nm,
corresponding to the aperture of our optical setup.
We exclude the first three Zernike polynomials in the optimization process,
starting from the radial degree $2$.
Indeed, the initial one, known as the piston, does not influence the PSF quality.
The subsequent ones, the tip and the tilt, cause the PSF to shift.
Our procedure relies on optimizing the maximum of the PSF, wherever that is,
rendering us indifferent to these two parameters.
After optimization,
it is possible to generate the correction pattern using a selected number of Zernike polynomials
in order to investigate their impact on the PSF quality.
We see here that using about 10 Zernike polynomials is sufficient to obtain a Strehl ratio $>0.99$.
Fig.~\ref{fig:zernike} demonstrates the Strehl ratio and the intensity profiles of the PSF
for different counts of the utilized Zernike polynomials.
It is important to note that we do not use the full surface of the DMD.
Using a larger area may lead to stronger deformations of the PSF,
requiring a larger number of Zernike polynomials to be corrected accurately.
The experimental data, in addition to the Python code used to generate the figures, can be accessed in the dedicated repository~\cite{github}.\\

\subsection{Python code example}

We provide in the paper repository~\cite{github} a Python code example
to simulate the effect of aberration on a DMD
and then perform a sequential optimization as previously proposed
to learn the characterize pattern.
We make use of the !aotools! package~\cite{Townson2019aotools} for generating Zernike polynomials.


\begin{tldr}
  \textbf{TL;DR:}\\
  DMDs are not flat and can introduce aberrations;
  these are typically much stronger than those commonly observed with liquid crystal SLMs.
  This can be counteracted in situ using a standard setup and a straightforward optimization procedure
  to maximize the intensity at the central position of the PSF.
\end{tldr}


\section{Mechanical and thermal stability}

Unlike the original purpose of the DMD, i.e. amplitude modulation for video projectors,
typical scientific applications require a high stability of the generated wavefront.
This is particularly true for applications in complex media,
such as strongly scattering media or multimode fiber,
where a small change in the phase front can lead to a large change in the output intensity profile.
While LC-SLMs design has been improved and adapted to scientific applications
over the last decades,
DMD are still relatively new tools for wavefront shaping and sensing
and are prone to instabilities that need to be addressed by the user.
In this section, we present the effect of mechanical and thermal instabilities
and how to limit their impact on the wavefront quality with simple solutions.\\

\subsection{Mechanical stability}

Most DMD kits consist of two primary components,
the chip itself and the electronic board that controls it.
This could be the standard electronics board typically used for video projectors,
as seen in TI evaluation kits,
or an FPGA specifically designed for rapid scientific usage,
as offered by Vialux~\cite{vialux} for instance.
Integral to these electronics is a fan designed to cool both the chip and the electronic board.
However, due to the use of a rigid flat cable for connection between the chip and the electronics board,
these parts are not mechanically independent.
As such, vibrations originating from the board are partially transmitted to the chip,
resulting in minute rotations of the mirror surface.
Although this perturbation is inconsequential for video projection,
they can have significant impacts on applications involving complex media
given their high sensitivity to phase front variations.\\

\begin{figure}[ht]
  \centering
  \includegraphics[width = 0.95\textwidth]{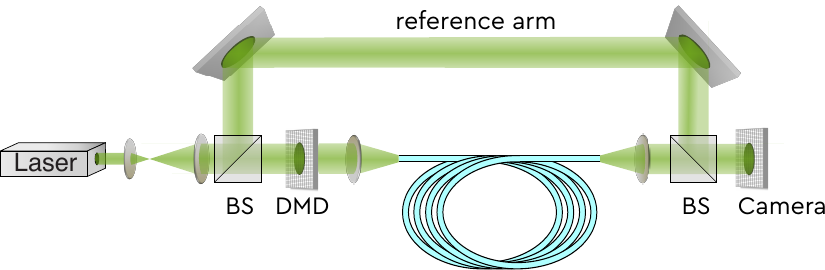}
  \caption{
    \textbf{Measuring stability at the output of a multimode fiber.}
    Schematic representation of the setup used to measure the phase fluctuations
    through a multimode fiber.
    An equivalent setup can be used with a scattering medium.
  }
  \label{fig:MMF_ref}
\end{figure}

Due to its high sensitivity in complex media,
it is convenient to characterize this effect
directly on the system's response,
rather than constructing a distinct setup to analyze the wavefront itself.
An example of such a setup is demonstrated in Fig.~\ref{fig:MMF_ref},
although a similar approach can be employed with a scattering medium.
We enlarge a laser beam onto the DMD and transmit the incoming light through a multimode fiber.
Additional elements are required in the setup to fulfill the requirement for complex modulation~\cite{Gutierrez2024DMD}.
For the sake of clarity, we present a simplified version of the setup where those elements are not present.
The output from the fiber is then made to interfere with a reference arm
in an off-axis configuration~\cite{Cuche2000spatial},
allowing us to detect changes in the output complex field
by recording the interference pattern using a camera.\\

\begin{figure}[ht]
  \centering
  \includegraphics[width = 0.75\textwidth]{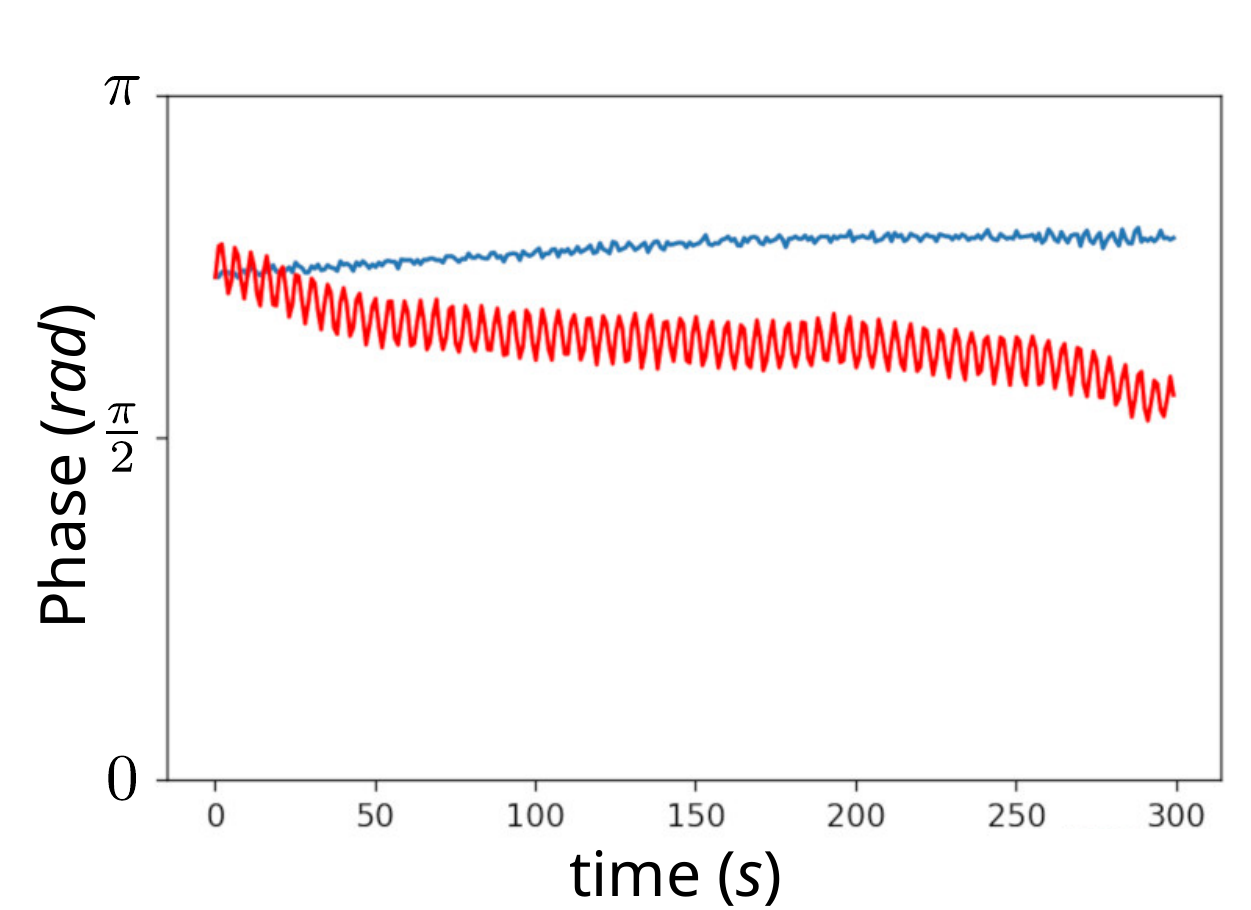}
  \caption{
    \textbf{Vibration induced phase fluctuations.}
    Phase fluctuations measured over time at
    a specific position on a plane located at the distal end of a multimode fiber.
    We employ off-axis holography using the setup depicted in Fig.~\ref{fig:MMF_ref}.
    The red and blue curves correspond to the phase fluctuations
    measured respectively without and with vibration damping
    by securing the flat cable with foam, as illustrated in Fig.~\ref{fig:dumping}.
  }
  \label{fig:phase_vibrations}
\end{figure}

In the supplementary materials~\cite{SI}, we present an animation illustrating the dynamic pattern.
In off-axis holography,
the local transverse displacement of the fringes
is directly proportional to the phase,
with a displacement equivalent to the period of the fringes corresponding to $2\pi$~\cite{Cuche2000spatial}.
This permits us to estimate the fluctuation in phase over time at a given position of the output plane.
As illustrated in Fig.~\ref{fig:phase_vibrations} (depicted by the red curve),
the phase varies rapidly over time,
a fluctuation attributed to the rapid mechanical vibrations transmitted by the board.\\

\begin{figure}[ht]
  \centering
  \includegraphics[width = 0.95\textwidth]{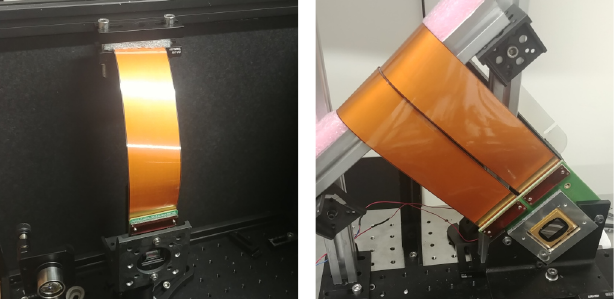}
  \caption{
    \textbf{\red{Damping} mechanical vibrations.}
    Pictures of experimental setups where a damping of the vibrations is implemented.
    (a) The flat cable is secured with foam and metallic plates.
    (b) The flat cable is in tension against a foam material supported by a metallic plate.
  }
  \label{fig:dumping}
\end{figure}

A simple yet effective solution consists in damping the vibrations
at the flat cable's level by clamping it with a soft material,
as depicted in Fig.~\ref{fig:dumping}.
This can be achieved using commonly available materials.
In this context, we utilize simple foam, typically used for packaging, and secure it to the cable
with two metallic plates, screws, and nuts.
We observe a significant decrease in the phase fluctuations,
as demonstrated in Fig.~\ref{fig:phase_vibrations} (blue curve).\\

\begin{tldr}
  \textbf{TL;DR:}\\
  The functioning of DMDs can be perturbed by vibrations transmitted from the electronics board
  via the rigid flat cable that links it to the DMD chip.
  This adverse effect can be minimized by securing the cable with a soft material
  that serves to dampen these vibrations.
\end{tldr}

\subsection{Thermal stability}

Electronics utilized to control the DMD chip experience thermal variations during operation.
The dynamics of this effect are dependent on the frame rate.
Specifically, the chips heats up more quickly when increasing the frame rate.
The increase of temperature can reach more than $15$ degrees Celsius
when running a sequence at maximum speed (20-30 kHz)~\cite{Rudolf2021thermal}.
Notably, this effect is less pronounced when the device is on but not running a sequence.
Temperature fluctuations can cause deformations on the chip's surface
and can modify the phase response of the glass protective window.
This creates low order aberrations, which degrade the quality of the wavefront.
While this issue is comparatively less critical than static aberrations and mechanical instabilities
previously detailed,
it nonetheless has a substantial impact on the complex medium's response
when a DMD is employed to modulate the input wavefront.\\

\red{Empirically, we observe deformations of the pattern 
at the output of the medium under study when the DMD chip heats up after switching on the device. 
These deformations originate from a temperature-dependent perturbation of the modulated field. 
We attribute this effect to the transparent part of the DMD 
(i.e., the protection glass window) 
having a refractive index that significantly changes with temperature. 
Once the device reaches thermal equilibrium, this effect stabilizes, 
allowing us to find the appropriate correction mask as described in Section 3 of the manuscript. 
It is important to stress that it is the thermal fluctuations that impact the stability of the modulation, 
not the temperature itself, 
as we obtain stable experimental conditions even when equilibrium is reached 
at the maximal operating temperature of the modulator without thermal cooling.\\}

\begin{figure}[ht]
  \centering
  \includegraphics[width = 0.95\textwidth]{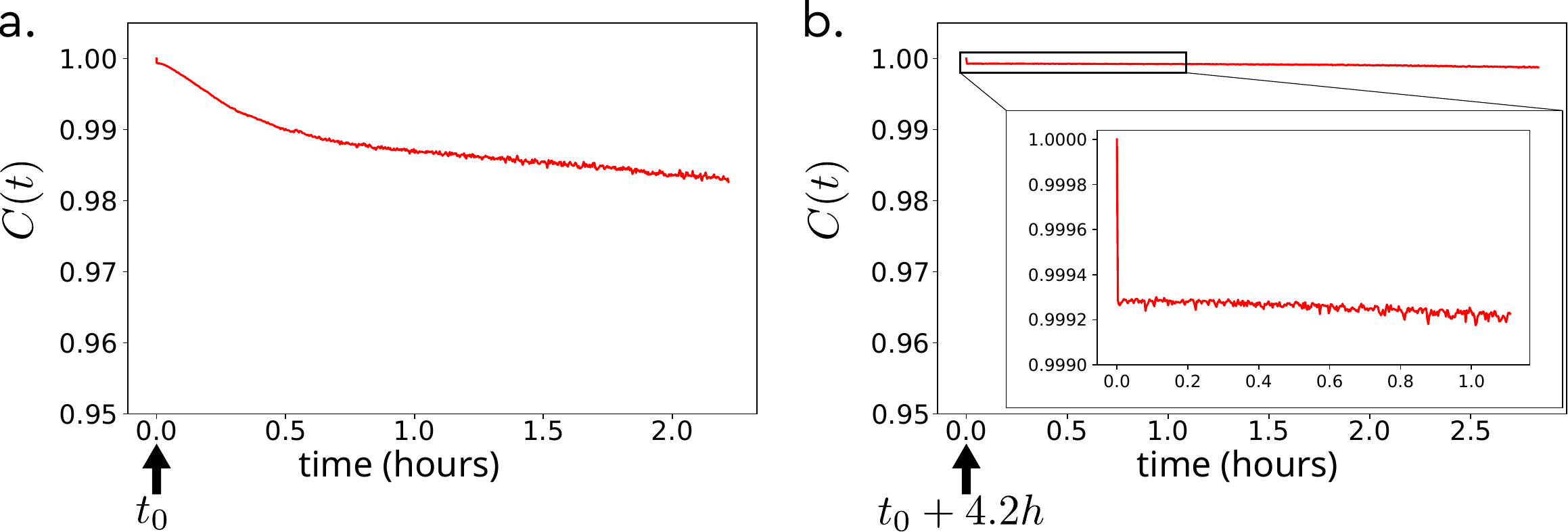}
  \caption{
    \textbf{Effect of temperature.}
    Temporal decorrelation of a speckle pattern measured at the output of a multimode fiber
    when running a sequence in an infinite loop.
    We use Eq.(\ref{eq:temp_decorr}) to estimate the correlation between the intensity pattern
    at a given time $t$ and the one at $t=0$.
    $t=0$ corresponds respectively to the start of the sequence (a)
    and to 4.2 hours after the start of the sequence (b).
  }
  \label{fig:temp_decorr}
\end{figure}

Before initiating a wavefront shaping experiment,
it is important to characterize the influence of temperature
to assess the extent to which it impacts the results.
Although the exact effect on the wavefront distortion can be directly quantified~\cite{Rudolf2021thermal},
it is typically more convenient
to directly measure the effect on the studied system's response. 
\red{Directly using the output of the complex medium under study presents multiple advantages. First, it allows the setup to remain unchanged and facilitates easy re-characterization of the effect if some parameters change. Additionally, complex media tend to be very sensitive to input phase changes, with the sensitivity varying from one medium to another. By using the complex medium itself, we obtain the exact sensitivity needed for the current setup. Therefore, regardless of the qualitative changes in the input field, stability in the output indicates that the setup is adequate for studying the medium in question.}
To do so, we use a setup similar to the one presented in the previous section
and depicted in Fig.~\ref{fig:MMF_ref}.
We can then estimate the field or intensity decorrelation over time.
We present here results with intensity correlation, as it does not require a reference arm.
The measured output pattern typically take the form of a seemingly random speckle pattern,
that is sensitive to minute changes in the input wavefront.
The correlation estimation is obtained by comparing the output intensity pattern
$I(\vec{r}, t)$ at a given time $t$
to the one at $t=0$.
We use the following expression for the correlation:

\begin{equation}
  C(t) =
  \frac{
    \left\langle
    \bar{I}(\vec{r}, t) \bar{I}(\vec{r}, t=0)
    \right\rangle_{\vec{r}}
  }{
    \sqrt{
      \left\langle
      \bar{I}(\vec{r}, t)^2
      \right\rangle_{\vec{r}}
      \left\langle
      \bar{I}(\vec{r}, t=0)^2
      \right\rangle_{\vec{r}}
    }
  } \, ,
  \label{eq:temp_decorr}
\end{equation}

with $\bar{I}(\vec{r}, t) = I(\vec{r}, t) - \left\langle I(\vec{r}, t) \right\rangle_{t}$
and $\left\langle . \right\rangle_{\vec{r}}$.
$\left\langle . \right\rangle_{t}$
represent the spatial averaging over the region of interest
of the output plane (i.e. the plane of the camera sensor)
and the temporal averaging over the measured frames.
Figure~\ref{fig:temp_decorr} shows the measured decorrelation
over time when running an sequence in an infinite loop.
$t=0$ correspond in the first case (a)
to the start of a sequence
\red{and in the second case (b) is set to
4.2 hours after the start of the sequence.}
For the scenario (a), it is worth noting that before the sequence commenced,
the DMD was active but remained in the idle state,
meaning that it was not executing a sequence.
We observe a non-negligible decorrelation over time
after the sequence started
that slows down after about 1 hour.
When running the same experiment 4.2 hours after starting the sequence,
the correlation is now stable.\\

\begin{figure}
  \centering
  \includegraphics[width = 0.65\textwidth]{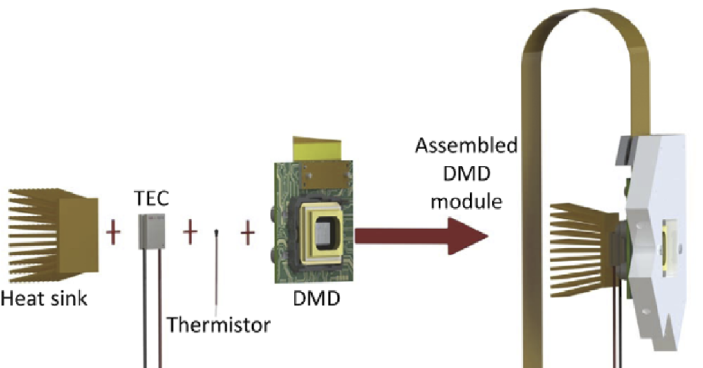}
  \caption{
    \textbf{Thermal stabilization of the DMD chip.}
    Schematic representation of the setup utilized to regulate the temperature of the DMD chip.
    A thermoelectric cooler and a heat sink is employed to control the temperature,
    while a thermistor is used to measure deviations from the temperature setpoint.
    A closed-loop system is used to regulate the temperature.
    Image adapted from~\cite{Rudolf2021thermal}.
  }
  \label{fig:decorr_T}
\end{figure}

The more efficient way to counteract this effect is to use a
closed-loop system to stabilize the temperature of the DMD chip.
This can be achieved by using a thermoelectric cooler
as demonstrated in Ref.~\cite{Rudolf2021thermal}
and depicted in Fig.~\ref{fig:decorr_T}.

\begin{tldr}
  \textbf{TL;DR:}\\
  DMDs take about an hour to thermally stabilize when a sequence is running.
  It can be countered by using a thermoelectric cooler
  and a temperature sensor.
  It can also be simply mitigated by letting a sequence run for few hours
  before starting the experiment.
\end{tldr}

\section{Conclusion}
DMDs are powerful tools for wavefront shaping applications in complex media
due to their high pixel count, relative low cost, and high refresh rate.
However, mostly due to their original purpose of amplitude modulation
of incoherent light for video projection,
several effects need to be taken into account when using DMDs for wavefront shaping.
First, the choice of its pixel pitch must be made carefully
by considering the wavelength of operation
to ensure a good modulation quality and diffraction efficiency.
\red{Furthermore}, the DMD surface is not flat and can introduce aberrations
which can be corrected using a simple optimization procedure.
Finally, the DMD is sensitive to mechanical vibrations and thermal variations,
which can be mitigated ensuring a good mechanical and thermal isolation of the device.


\section*{Data and code availability}

\noindent Data and code examples are available in the dedicated repository~\cite{github}.

\section*{Appendix A: 1D calculation of the DMD diffraction effect}
\label{app:1D}

Assuming the effect of the device's finite size and illumination to be negligible,
we can represent the field reflected from the device
under the influence of plane wave illumination
across two systems as follows.
Here, $\theta_B$ denotes the blaze angle, i.e. the tilt of the micromirror normal
with respect to the device normal, so that the corresponding specular direction is
$2\theta_B-\alpha$:

\begin{equation}
  \begin{aligned}
    R_\text{flat}(x)   & \propto
    \left[\Pi\left(x/d'\right) \otimes_x \sum_k \delta(x-k d)\right]
    e^{j\frac{2\pi}{\lambda}\sin(\alpha) x} \,,\\
    R_\text{blazed}(x) & \propto
    \underbrace{
      \left(
        \underbrace{\Pi\left(x/d'\right)}_\text{pixel size}
        \,
        \underbrace{e^{j \frac{2\pi}{\lambda}\left( \sin(2\theta_B-\alpha) - \sin(\alpha) \right) x}}_{\substack{\text{blazed angle}  \, , \\ \text{+ angle of incidence}}}
      \right)
    }_\text{pixel response}
    \otimes_x
    \underbrace{\sum_k \delta(x-k d) }_\text{periodicity}
    \,
    e^{j\frac{2\pi}{\lambda}\sin(\alpha) x}
    \, .
  \end{aligned}
  \label{eq:grating_response}
\end{equation}

with $\Pi(x)$ the rectangular function,
representing the finite size of the pixel,
defined as:

\begin{equation}
  \Pi(x) =
  \begin{cases}
    1, & \text{if } -\frac{1}{2} < x < \frac{1}{2}, \\
    0, & \text{otherwise}.
  \end{cases}
\end{equation}

The filling fraction, in the 1D case, is the ratio of the size of the pixels (or micro-mirrors) $d'$,
and the pitch $d$, such that $d'=\rho d$.\\

The Fourier transform of Eq.~\ref{eq:grating_response} can be written as:

\begin{equation}
  \begin{aligned}
    I_\text{flat}(\theta) \propto
      \sum_p \delta\bigl(\sin(\theta)+\sin(\alpha)-p\,\sin(\theta_D)\bigr)
    \times
    \operatorname{sinc}^2\left( \pi \rho\frac{
        \sin(\theta)-\sin(\alpha)
        }{
        \sin(\theta_D)
        }\right) \\
    I_\text{blazed}(\theta) \propto
    \underbrace{
      \sum_p \delta\bigl(\sin(\theta)+\sin(\alpha)-p\,\sin(\theta_D)\bigr)
    }_\text{orders of diffraction}
    \times
    \underbrace{
      \operatorname{sinc}^2\left( \pi \rho\frac{
        \sin(\theta)-\sin(2\theta_B - \alpha)
        }{
        \sin(\theta_D)
        }\right)
    }_\text{envelope} \, .
  \end{aligned}
\end{equation}

We observe that the envelope (right-hand term) is maximal for
$\sin(\theta_\text{max}) = \sin(2\theta_B - \alpha)$,
while the effect of the periodicity (left-hand term) is maximal when
$\sin(\theta_p) + \sin(\alpha) = p\,\lambda/d$,
representing the orders of diffraction.


\section*{Appendix B: 2D calculation of blazed grating condition}

To analyze more precisely the effect of diffraction in a DMD,
one needs to consider the 2D surface of the modulator.
We can establish a Cartesian coordinate system on the plane of the DMD,
with axes $x$ and $y$ aligned with the pixel sides
(refer to Fig.~\ref{fig:2d_geom}a).
The pixels are uniformly repeated along these axes.
However, a technical challenge arises
in that the axis of rotation of the pixels
aligns with the pixel diagonals,
resulting in a rotation by 45 degrees with respect to the $x$ and $y$ axes.
For the convenience of alignment and manipulation of the optical setup,
it is preferable to work with the incident and outgoing beams
which have the optical axis contained in the horizontal plane,
i.e. a plane parallel to the table surface.
A straightforward and common solution is to rotate the chip by 45 degrees
relative to the horizontal plane,
which aligns the pixel axis of rotation to be vertical.

A more detailed description of the 2D system 
and computation of its response 
can be found in~\cite{Scholes2019structured}.
However, we can find a simple condition for
having most of the energy in one diffraction order 
by looking at the effect of the tilts of the mirrors 
when the axis is rotated by 45 degrees compared to 
the axis of the pixel array (axes $x$ and $y$). 
We consider a plane wave with an incoming angle $\alpha$ contained in the horizontal plane 
(i.e. orthogonal to the rotation axis of the micro-mirrors).
The effect of this angle is represented by a phase term of the form:

\begin{equation}
{\vec{k}_{in}\cdot\vec{r}} = {\frac{2\pi}{\lambda}\sin\alpha \frac{\left(x - y\right)}{\sqrt{2}}}
\, ,
\end{equation}

with $\vec{k}_{in}$ the incoming wavevector and $\vec{r}$ the position in the modulator plane, 
represented by the coordinate $(x,y)$.

The grating condition is achieved when the output contributions are in phase 
for the output angle corresponding to the specular reflection on the tilted mirrors. 
It corresponds to the plot on the left in Fig.\ref{fig:gratings} for the 1D case.
The phase contribution resulting from an output angle $2\theta_B-\alpha$, 
corresponding to the specular reflection on each micro-mirror,
reads:

\begin{equation}
\vec{k}_{out}\cdot\vec{r} 
= {\frac{2\pi}{\lambda}\sin\left(2\theta_B - \alpha\right) \frac{\left(x - y\right)}{\sqrt{2}}}
\, ,
\end{equation}

The total phase then reads:

\begin{equation}
\phi(x,y) = \vec{k}_{in}.\vec{r}  + \vec{k}_{out}.\vec{r} =  
{\frac{2\pi}{\lambda}\sin\left(2\theta - \alpha\right) \frac{\left(x - y\right)}{\sqrt{2}}} +
{\frac{2\pi}{\lambda}\sin\alpha \frac{\left(x - y\right)}{\sqrt{2}}} 
\, .
\end{equation}

We see that $x$ and $y$ have the same effect.
The grating condition is then achieved when two consecutive pixels, 
separated by a distance $d$ in both the $x$ and $y$ directions, are in phase.
I.e. for

\begin{equation}
  \phi(x,y) = \phi(x+d,y) = \phi(x,y+d) \, .
\end{equation}

The new blazed grating equation then reads~:

\begin{equation}
  \text{sin}(2\theta_B-\alpha) + \text{sin}(\alpha)
  = 2 \,\text{sin}(\theta_B)  \,\text{cos}(\theta_B-\alpha)
  = p\frac{\sqrt{2}\lambda}{d} \, ,
  \label{eq:blazed_eq_new}
\end{equation}

\section*{Acknowledgments}

\noindent 
S.M.P, M.W.M, and R.G.C acknowledge the French \textit{Agence Nationale pour la Recherche} grants No. ANR-23-CE42-0010-01 MUFFIN and No. ANR-20-CE24-0016 MUPHTA 
and the  Labex WIFI grant No. ANR-10-LABX-24, ANR-10-IDEX-0001-02 PSL*. 

L.M, J.C, M.G, and R.G acknowledge Cathie Ventalon for sharing a V-module and for her valuable discussions. L.M, J.C, M.G, and R.G were funded by Centre National de la Recherche Scientifique MITI/80Prime and the iXcore - iXlife - iXblue Fundation.

\end{document}